\documentclass{elsart}
\usepackage{graphicx}
\usepackage{amssymb}
\begin{document}

\begin{frontmatter}

\title{Using Fractal Dimensionality in the Search for Source Models of Ultra-High Energy Cosmic Rays} 

\author{B.T.~Stokes\corauthref{cor1},}
\author{C.C.H.~Jui,} 
\author{and J.N.~Matthews}
\address{University of Utah,
Department of Physics and High Energy Astrophysics Institute,
Salt Lake City, Utah, USA}
\corauth[cor1]{
Corresponding~author.\ {\it E-mail~address}:~stokes@cosmic.utah.edu~(B.T.~Stokes)
}

\begin{abstract}
Although the existence of cosmic rays with energies extending well 
above $10^{19}$~eV has been confirmed, their origin remains one of the most 
important questions in astro-particle physics today.  Several source 
models have been proposed for the observed set of Ultra High Energy 
Cosmic Rays (UHECRs).  Yet none of these models have been conclusively 
identified as corresponding with all of the available data.  
One possible way of achieving a global test
of anisotropy is through the measurement of the information dimension, 
$D_{\rm I}$, of the 
arrival directions of a sample of events.  $D_{\rm I}$ is a measure of the 
intrinsic heterogeneity of a data sample.  We will show how this method can 
be used to take into account the extreme asymmetric angular resolution 
and the highly irregular aperture of 
a monocular air-fluorescence detector.  We will then use a simulated, 
isotropic event sample to show how this method can be used to place upper
limits on any number of source models with {\it no} statistical penalty. 
\end{abstract}

\begin{keyword}
cosmic rays \sep anisotropy \sep fractal dimensionality \sep Centaurus A \sep
dark matter halo \sep air fluorescence

\PACS 98.70.Sa \sep 95.55.Vj \sep 95.85.Ry \sep 95.35.+d

\end{keyword}

\end{frontmatter}

\section{Introduction}

The observation of Ultra High Energy Cosmic Rays (UHECRs) has now spanned 
nearly three decades.  Over that period, 
many different source models have been 
proposed to explain the origin of these remarkable events.  Recently, the
Akeno Giant Air Shower Array (AGASA) reported clustering at small
angular scales for the events that were observed above $4\times10^{19}$~eV 
\cite{Takeda:1999sg}.  However, this result could not be 
confirmed by the High Resolution Fly's Eye (HiRes) air fluorescence detector 
\cite{bellido} despite the fact that Hires-I's monocular exposure was more 
than twice that of AGASA within the pertinent energy range \cite{jui}.  AGASA
has also reported a possible excess of events in the direction of the 
galactic center for events with energies around $10^{18}$~eV \cite{hayashida}.
However, HiRes reported that it did
{\it not} see anisotropies when examining harmonics in right ascension, 
{\it a priori} determined point sources, or enhancement in the supergalactic
plane.  Furthermore, prior analysis by 
the the original Fly's Eye showed no evidence of anisotropies when dependencies
in galactic latitude and longitude, harmonics in right ascension, excess maps,
and specific {\it a priori} determined point 
sources were examined \cite{bird:1,bird:2}.  However, 
Fly's Eye did find
a small apparent excess in the direction of the galactic plane.
\cite{bird:gal}.
In 1995, it was reported by Stanev {\it et al.} 
that the combined data of 
Haverah Park, Yakutsk, and AGASA showed an excess along the supergalactic 
plane with several potential point sources for events above 
$2\times10^{19}$~eV \cite{stanev}.

These conflicting reports call for developing a more global way in which 
one could determine if a given sample possesses
any statistically significant anisotropy.  We will show that by 
considering the information dimension, $D_{\rm I}$, of a given sample, 
one can simultaneously 
look for anisotropies at all angular scales greater than the angular 
resolution of the sample by considering the intrinsic heterogeneity of that 
particular data sample.  This method is quite robust in that it can
easily accommodate both asymmetric angular resolutions and irregular apertures.
Furthermore, in the event that a sample is shown to be consistent with an 
isotropic distribution, this method can be used to place upper limits
on possible source models.  Because only a single measurement
is taken of the actual data, any number of source models can be considered 
simultaneously without incurring statistical penalties. 

\section{Calculating Fractal Dimensions of a Data Sample}

Fractal dimensionality is a simple measure of the scaling 
symmetry of a structure.  
By measuring the fractal dimension of a data sample, one can examine
its self-consistency at different levels of magnification.  There are several
ways of exploiting this idea.  
From a computational perspective, the simplest is to use 
box-counting.  For the most general case, the 
capacity dimension, $D_{\rm c}$ \cite{kolmogorov}, one partitions 
the sample space into equi-sized and equi-shaped ``boxes'' 
with edge size $\epsilon$:
\begin{equation}
D_{\rm c}=\lim_{\epsilon \to 0^{+}} \frac{\log{\it N}(\epsilon)}{\log1/\epsilon}.
\end{equation}
Here, $N(\epsilon)$ is the minimum number
of ``boxes'' with edge size $\epsilon$ necessary to cover one's
sample.  

The capacity dimension has a serious limitation:  It only looks
for the presence of the sample within the available space and does not 
consider variations in the density of the sample at a given point in space. 
In cases where the density may differ within the sample space, the appropriate
alternative is to use the
information dimension, $D_{\rm I}$ \cite{balatoni,nayfeh}:  
\begin{equation}
D_{\rm I}=-\sum_{i=1}^{N}\lim_{\epsilon \to 0^{+}} 
\frac{P_{\rm i}(\epsilon)\log P_{\rm i}(\epsilon)}{\log 1/\epsilon},
\end{equation}
where $P_{\rm i}(\epsilon)$ 
is the probability of finding a data point in the $i$-th
box of edge size $\epsilon$.  This is a particularly suitable
measurement when considering a data set consisting of UHECR arrival 
directions with finite angular resolution.

It should be noted that $D_{\rm C}$ and $D_{\rm I}$ are both particular cases
of the $q$-dimension \cite{grassberger,hentschel},
\begin{equation}
D_q=\frac{1}{1-q}\lim_{\epsilon \to 0^{+}}\frac{\log I(q,\epsilon)}
{\log 1/\epsilon};
\end{equation}
where
\begin{equation}
I(q,\epsilon)=\sum_{i=1}^{N}P_{\rm i}(\epsilon)^q.
\end{equation}
We can then see that $D_{\rm C}=\lim_{q \to 0}D_q$ and that 
$D_{\rm I}=\lim_{q \to 1}D_q$.

\section{Application to Arrival Direction Distributions for UHECRs}

In principle, it is simple to calculate the information dimension for a given
sample of events.  However, two complications arise when considering a
set of arrival directions of UHECRs.  First, the event directions 
are not known with complete precision.  This makes the determination of
$D_{\rm I}$ as $\epsilon \to 0$ meaningless.  Secondly, the determination
of $D_{\rm I}$ requires that the sample space be divided into equi-sized and 
equi-shaped bins.  For a spherical surface, this is simply not
possible.  Nevertheless, 
there are workable solutions for both of these problems.

\subsection{Angular Resolution}

We will consider a hypothetical monocular air-fluorescence detector.
This detector is largely based upon the characteristics of the 
Hires-I detector \cite{Abu-Zayyad:2002ta,Abu-Zayyad:2002sf}.
Our detector observes events with an angular resolution that is described by
a highly asymmetric 2-d Gaussian.  
\begin{figure}[t,b]
\begin{center}
\begin{tabular}{c}
\includegraphics[width=7.0cm]{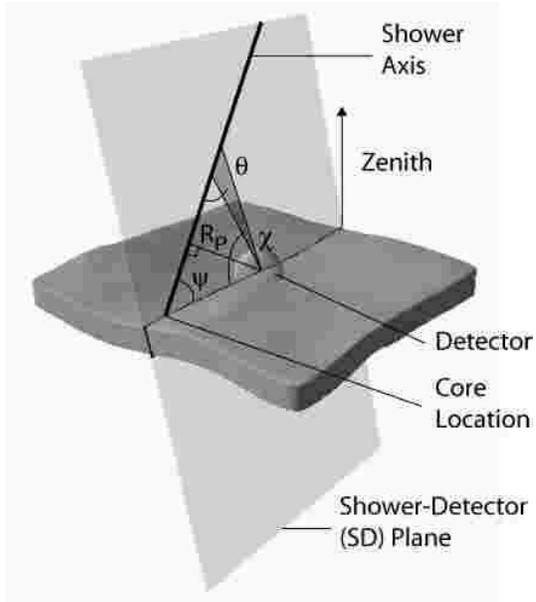}\\
\end{tabular}
\end{center}
\caption{The geometry of reconstruction for a monocular air fluorescence 
detector}
\label{figure:picture}
\end{figure}
For a monocular air fluorescence detector, angular resolution consists
of two components: $\sigma_{\rm 1}$,  
in the determination of the angle, $\psi$, within
the plane of reconstruction, and $\sigma_{\rm 2}$,  
in the estimation of the plane of 
reconstruction itself.  Figure~\ref{figure:picture} illustrates how
this geometry would appear with a particular plane of reconstruction and a 
particular value for $\psi$.  Intuitively, we can see that we should be able
to determine the plane of reconstruction quite accurately.  However, the value 
of $\psi$ is more difficult to determine because it is
dependent on the precise results of the monocular reconstruction 
\cite{Abu-Zayyad:2002ta,Abu-Zayyad:2002sf}.

The actual parameterizations of $\sigma_{\rm 1}$ and $\sigma_{\rm 2}$ assumed
are as follows:
\begin{equation}
\sigma_{\rm 1}=20^\circ e^{-1.5\log_{\rm 10}E_{\rm EeV}} + 4^\circ 
\label{equation:sigma1}
\end{equation}
and
\begin{equation}
\sigma_{\rm 2}=100^\circ e^{-0.5\Delta\chi} + 0.4^\circ. 
\label{equation:sigma2}
\end{equation}
Here, $E_{\rm EeV}$ is the primary energy of the shower in 
EeV. For the purpose of this study, the energy will be allowed to vary between 
$10^{18.5}$~eV and $10^{20}$~eV with a differential spectral index of $-2.7$.
In this scenario, a shower with a primary energy of $10^{18.5}$~eV 
will have $\sigma_{\rm 1}=13.4^\circ$, while 
a shower with a primary energy of $10^{20}$~eV will have 
$\sigma_{\rm 1}=5.0^\circ$. This difference can be attributed to the fact that
larger showers have better defined profiles and a better signal-to-noise 
ratio. 

The factor, $\Delta\chi$, in equation \ref{equation:sigma2} 
is the angular track length (in degrees) 
of the shower as observed by the detector, which is allowed to vary 
between $8^\circ$ and $30^\circ$.  A shower
with an observed track length of $8^\circ$ will have 
$\sigma_{\rm 2}=2.2^\circ$, while a shower with an observed track length of 
$30^\circ$ will have $\sigma_{\rm 2}=0.4^\circ$;  a longer track-length leads 
to a more accurate determination of the plane of reconstruction.  The 
distribution of $\sigma_{\rm 2}$ values is largely independent of energy 
because while higher energy showers do lead to more longitudinal development,
they are also brighter, which allows one to observe them at greater distances.
These competing factors lead to $\Delta\chi$ distributions that are virtually
identical across the observed spectrum.

In general, it should be noted the distributions of $\sigma_{\rm 1}$ and 
$\sigma_{\rm 2}$ values are relatively insensitive to the 
differential spectral index that 
is chosen.  We ascertained this by considering two sets, one with a 
differential spectral
index of $-2.5$ and one with a differential spectral index of $-3.5$.  
Even for a variation
that was much larger than the accepted range of experimental values for the
the UHECR spectrum 
\cite{Abu-Zayyad:2002ta,Abu-Zayyad:2002sf,Bird:wp,Takeda:1998ps}, the value
of $\bar{\sigma_{\rm 1}}$ increased by only $11\%$.  The value of 
$\bar{\sigma_{\rm 2}}$ remained unchanged.  This can be explained by 
realizing for a steeply falling spectrum, the overwhelming majority of 
observed events in either case will occur in the first half decade of the 
measurement.  This is a very small effect compared to the expected statistical
fluctuations that would occur between two consecutive sets of observations. 

For the purpose of calculating $D_{\rm I}$, we can treat the arrival 
direction of each individual shower as a two dimensional elliptical Gaussian 
distribution with the parameters $\sigma_{\rm 1},\sigma_{\rm 2}$.  
The size of a bin's edge, $\epsilon$, 
will be allowed to take on a series of values, 
$\Delta\theta$, which will be in an interval corresponding to a scale length 
of the sample or in the case of a smooth distribution, the smallest value
that is computationally feasible.  For finite events samples, we will use
$\Delta\theta\simeq0.5^\circ$.  In the case of smooth distributions we will
use a computationally limited value of $\Delta\theta=1/6^\circ$.
The number of points in each shower direction distribution, $N_{\rm Dist}$, 
will be determined by the mean value, $<\!\! n_{\rm i}\!\! >$, 
necessary to assure that the fractional Gaussian fluctuations of the count, 
$n_{\rm i}$, in each bin, do not on average, exceed a predetermined value 
(i.e. for $5\%$ fluctuations, $N_{\rm Dist}\simeq500$).  
For each value of $\epsilon$, the probability, $P_{\rm i}(\epsilon)$, 
for the $i$-th bin will be:
\begin{equation}
P_{\rm i}(\epsilon)=\frac{n_{\rm i}}{N_{\rm Dist}\cdot N_{\rm Shower}}.
\end{equation}
We calculate $D_{\rm I}$ for each value of $\epsilon$:
\begin{equation}
D_{\rm I}(\epsilon)=-\sum_{i=1}^{N} 
\frac{P_{\rm i}(\epsilon)\log P_{\rm i}(\epsilon)}{\log 1/\epsilon}.
\end{equation}
We then determine $D_{\rm I}$ to be 
$<\!\! D_{\rm I}(\epsilon)\!\! >$ over the specified interval of
$\epsilon$ values.

\subsection{Latitudinal Binning}

For the purpose of calculating $D_{\rm I}$, 
it is necessary that all bins be equi-sized
and equi-shaped as we vary the size of the side of the bins, $\epsilon$.  
While it is impossible to 
achieve completely this criterion on the surface of a sphere, we will be able
to do so approximately by adopting a latitudinal binning scheme.

Latitudinal binning is achieved by first dividing the sky into $N_{\rm \delta}$
declinational $(\delta)$ bands where each band has a width
\begin{equation}
\Delta\theta=\frac{\pi}{N_{\rm \delta}}
\label{equation:delthet}
\end{equation}
For each declinational band, the sky is then divided 
into  $N_{\rm\alpha,\delta}$ bins in right ascension $(\alpha)$ where:
\begin{equation}
N_{\rm \alpha,\delta}=\biggr[\frac{2\pi\int_{\delta 1}^{\delta 2} \cos\delta \, d\delta}{(\Delta\theta)^2}\biggr]=\biggr[\frac{2(N_{\rm\delta})^{2}\int_{\delta 1}^{\delta 2} \cos\delta \, d\delta}{\pi}\biggr].
\end{equation}   
The solid angle, $\Delta\Omega_{\rm\delta}$ of each bin (in steradians) is:
\begin{equation}
\Delta\Omega_{\rm \delta}=\frac{2\pi\int_{\delta 1}^{\delta 2} \cos\delta \, d\delta}{N_{\rm\alpha,\delta}},
\end{equation}
with a minimum value of $(\Delta\theta)^2$ (at the equator) and a 
maximum value of $\frac{\pi}{3}(\Delta\theta)^2$ (at the poles)
regardless of the value of $N_{\rm \delta}$.
This provides us with bins that are all almost the same area and nearly
 square-shaped (with the exception of three triangular bins at each pole).
The total number of bins in the sky can be approximated by:
\begin{equation}
N_{sky}\simeq4\pi(\frac{N_{\rm\delta}}{\pi})^{2}=
\frac{4}{\pi}(N_{\rm\delta})^{2}.
\label{equation:binnum}
\end{equation}

\begin{figure}[t,b]
\begin{center}
\begin{tabular}{c@{\hspace{0cm}}c}
(a)\includegraphics[width=6.15cm]{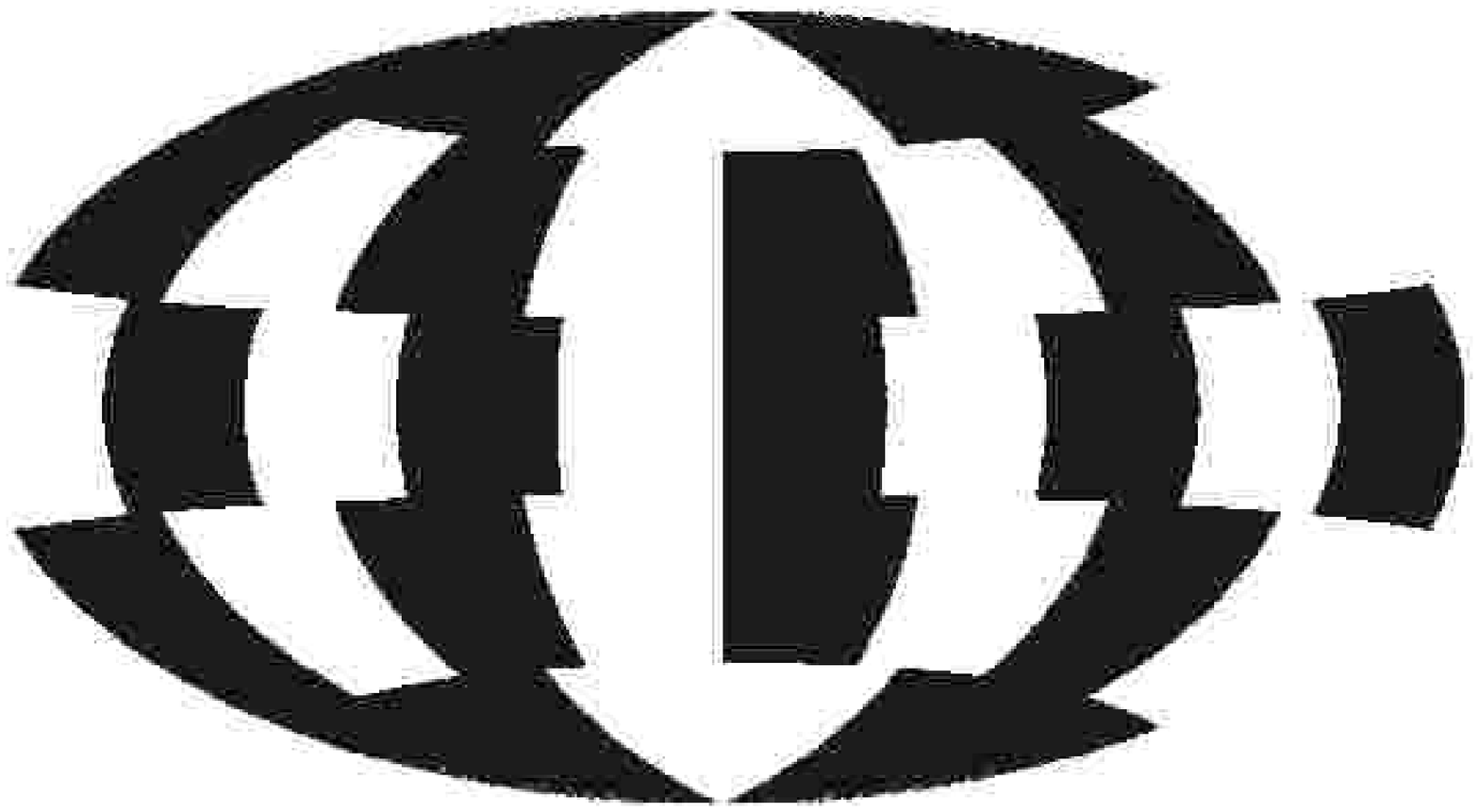}&
(b)\includegraphics[width=6.15cm]{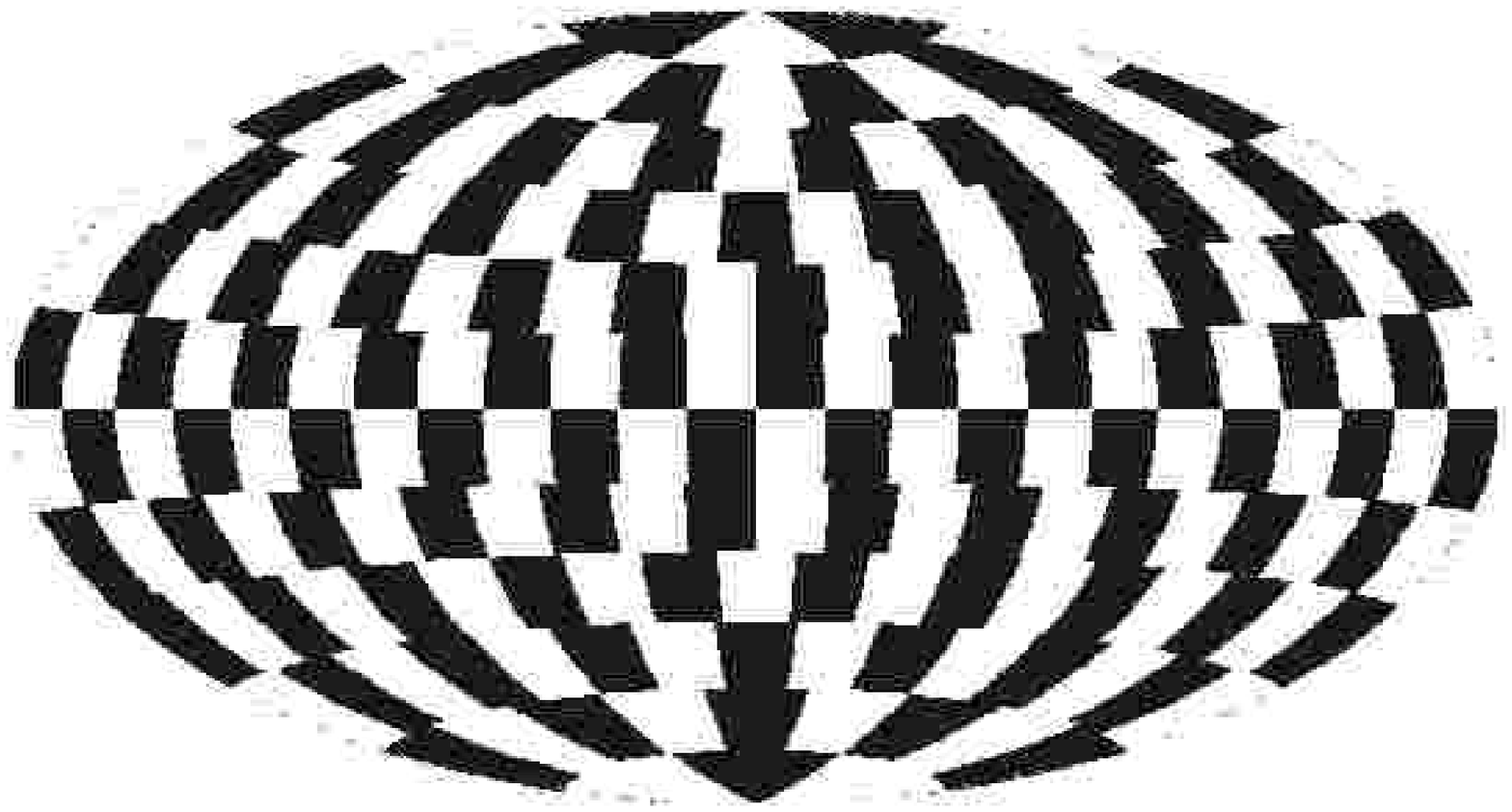}\\
(c)\includegraphics[width=6.15cm]{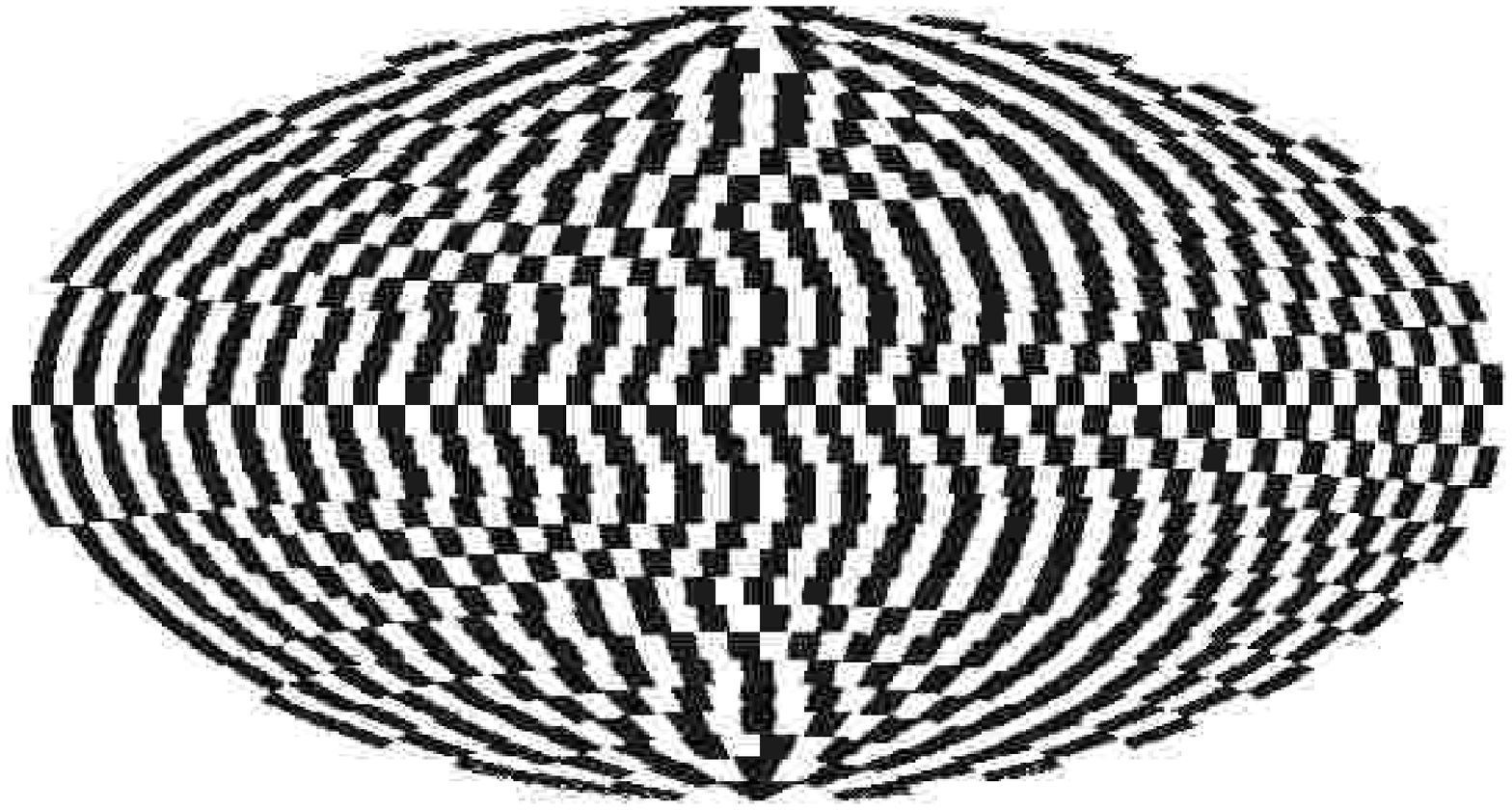}&
(d)\includegraphics[width=6.15cm]{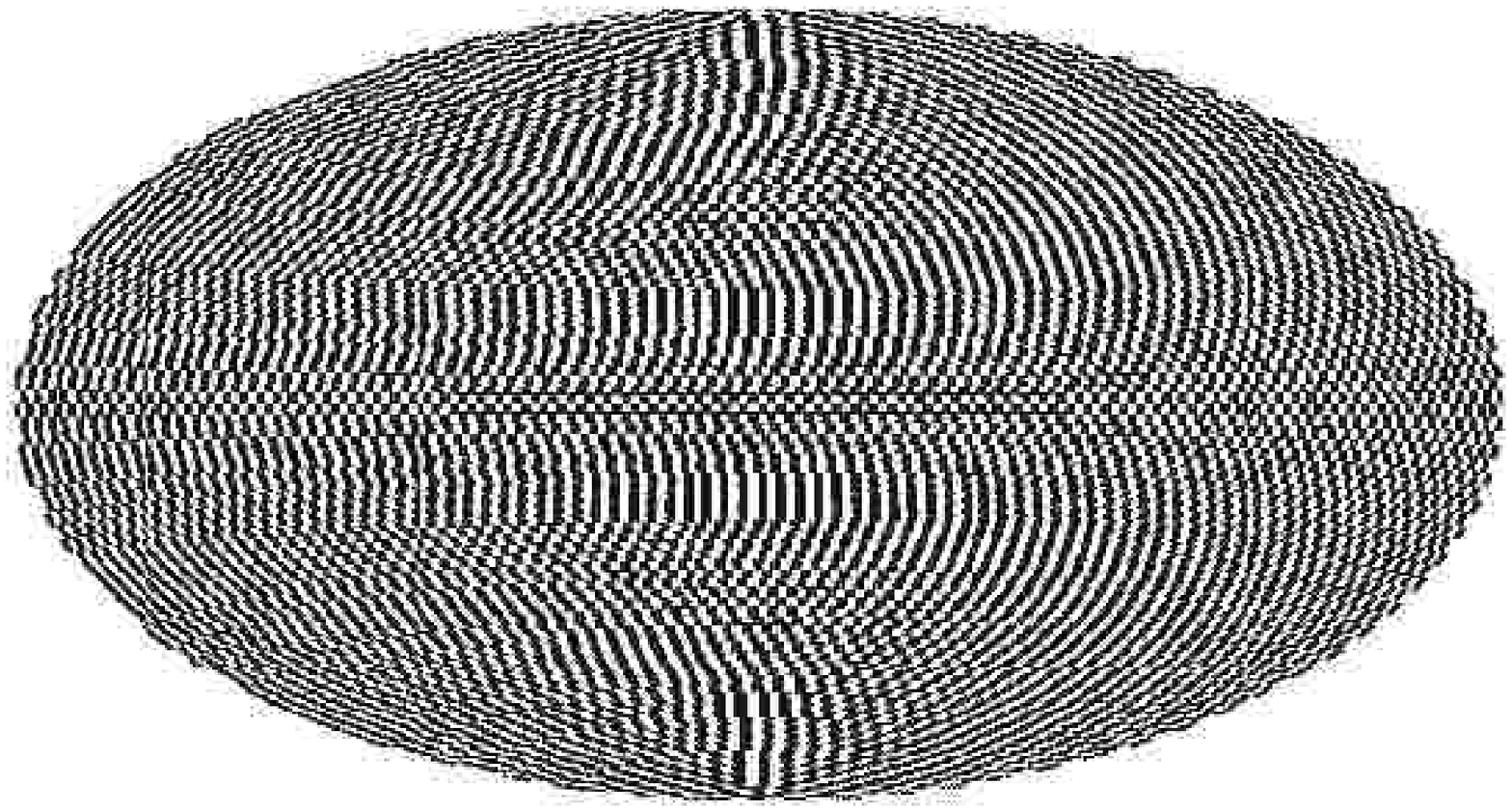}
\end{tabular}
\end{center}
\caption{Hammer-Aitoff projection of latitudinal bins for four different
values of $N_{\rm\delta}$---(a): $N_{\rm\delta}=5$; (b): $N_{\rm\delta}=12$; 
(c): $N_{\rm\delta}=30$; (d): $N_{\rm\delta}=90$. }
\label{figure:lat}
\end{figure}

In figure \ref{figure:lat}, we visualize the latitudinal binning technique 
for a series of different $N_{\rm\delta}$ values.

\subsection{Application to the Calculation of $D_{\rm I}$}

We can now apply the preceding machinery 
to the calculation of $D_{\rm I}$:  First,
we need to normalize the event count in each bin by its respective bin area, 
$\Delta\Omega_{\rm\delta}$: 
\begin{equation}  
P_{\rm i}(\epsilon)=\frac{n_{\rm i}(\Delta\theta)^2}
{N_{\rm Dist}N_{\rm Shower}\Delta\Omega_{\rm \delta}}=
\frac{n_{\rm i}\pi^{2}}{N_{\rm Dist}N_{\rm Shower}(N_{\rm\delta})^{2}
\Delta\Omega_{\rm\delta}}.
\label{equation:Pi1}
\end{equation}
If we then realize that $\epsilon=\frac{1}{N_{\rm \delta}}$, we can obtain:
\begin{equation}
D_{\rm I}(N_{\rm \delta})=-\frac{1}{\log N_{\rm \delta}}\sum_{i=1}^{N} 
P_{\rm i}(N_{\rm \delta})\log P_{\rm i}(N_{\rm \delta}).
\label{equation:Di1}
\end{equation}

This expression is reminiscent of the the general formula for entropy 
from statistical mechanics:
\begin{equation}
S=-k\sum_{r}p_{r}\log p_{r};
\end{equation}
where $p_r$ is the probability of a particle being the $r$-th state and $k$ is
the Boltzmann constant, which can be thought of as
a scaling constant based upon the intrinsic scale of the given particle. 
The information dimension, $D_{\rm I}$, is an analogous measurement of
the heterogeneity of a given data set.

\section{Calculating {\boldmath$D_{\rm I}$} for Exposures of Different Source Models}

\subsection{Exposure-Independent Source Descriptions}

We began by examining four different source models  
independently of detector exposure.  This allows us to calculate the value
of the information dimension, $D_{\rm I}$, 
without consideration to the detector aperture or statistical 
fluctuations from a finite event sample.  The source models are: 
an isotropic source model, a dipole source model, a model with
seven sources superimposed on an isotropic background, and a dark matter halo
source model. 

\subsubsection{Isotropic Model}

The first model that we will consider is an isotropic source model 
with distribution:
\begin{equation}
n_{\rm isotropic}=1.
\end{equation}

\subsubsection{Dipole Model}

The second is the Centaurus A dipole source model first proposed by
Farrar and Piran \cite{farrar}.  This model has a distribution of arrival 
directions characterized by a scaling parameter, $\alpha$, 
which can take on any
value between $-1$ and $+1$ and by $\theta$, which is the opening angle between
a given event arrival direction and the center of the dipole distribution at
Centaurus A.  The overall distribution is:
\begin{equation}
n_{\rm dipole}=1+\alpha\cos\theta.
\label{equation:dipole}
\end{equation}

\subsubsection{Discrete Source Model}

The third model that we will consider is one with seven discrete
sources superimposed on an isotropic background.  
For simplicity's sake, we will assume that all seven sources have an equal 
intensity indirectly determined by a parameter, $F_s$, which will be defined 
as the fraction of the entire event sample which originates in the seven 
sources.  
We will define our source direction to be the centroids of the seven 
hypothetical point sources proposed by the AGASA collaboration 
\cite{Takeda:1999sg}.  The equatorial coordinates used for
each point source are listed in table~\ref{table:7source}
\begin{table}[t,b]
\begin{center}
\begin{tabular}{|c|c|c|} \hline
Cluster & {\it Right Ascension} & {\it Declination} \\ \hline\hline
C1 & 01h13m & $20.6^\circ$ \\ \hline
C2 & 11h17m & $56.9^\circ$ \\ \hline
C3 & 18h51m & $48.2^\circ$ \\ \hline
C4 & 04h38m & $30.0^\circ$ \\ \hline
C5 & 16h02m & $23.3^\circ$ \\ \hline
C6 & 14h11m & $37.4^\circ$ \\ \hline
C7 & 03h03m & $55.5^\circ$ \\ \hline 
\end{tabular}
\\
\caption{Coordinates used for the centers of seven discrete sources.  These 
coordinates correspond to the centers of the seven clusters reported by the
AGASA Collaboration \cite{Takeda:1999sg}.}
\end{center}
\label{table:7source}
\end{table}

The arrival directions for the events from each source are  
assumed to be subjected to magnetic smearing.  That is, in the course of
traveling through space from the source to the point of observation, the
velocity vector of the event is subject to bending in the galactic and
extra-galactic magnetic fields.  We assumed that this bending produces 
an apparent source that can be characterized by a Gaussian distribution:
\begin{equation}
P(\Delta\theta)=
\frac{\Delta\theta}{\lambda^2}e^{-\frac{(\Delta\theta)^2}{2\lambda^2}},
\label{eqn:lambda}
\end{equation}
where $P(\Delta\theta)$ is the probability that an event will be observed with
an opening angle, $\Delta\theta$, from the nominal direction of the apparent 
source.  We will also assume that the arrival directions of the
events from all the sources are subject to the same degree of magnetic 
smearing, parameterized by $\lambda_\circ=1.5105\cdot\lambda$.  The parameter, 
$\lambda_\circ$, corresponds to the 68\% confindence interval in 
$\Delta\theta$.  For this paper, we will assume that $\lambda_\circ=5^\circ$.  
It should be noted that the nominal direction of the apparent source
is not necessarily the direction of the {\it actual} source because the
possibility exists that the path of the events in question traveled through
large regions of homogeneous magnetic fields.  

\subsubsection{Dark Matter Halo Model}

The fourth model that we will consider is  a dark matter halo source model.  
Dark matter halos are characterized by a density profile that is assumed to
take the Navarro-Frenk-White (NFW) form \cite{navarro}:
\begin{equation}
\rho_{\rm NFW}=\frac{\rho_\circ}{r(1+r/r_{\rm s})^2},
\label{equation:nfw}
\end{equation}
where $\rho_\circ$ is a dark matter
density parameter, $r$ is the distance from the center
of the halo, and $r_{\rm s}$ is a critical radius.  For our source model, we 
will consider the contribution of only the four closest significant dark matter
halos: the Milky Way, M31, LMC, and M33.  
We will assume that $\rho_\circ$ is the same for all four sources and 
that $r_{\rm s}$ scales with the cube root of the luminosity, 
$L^{\frac{1}{3}}$.  
Thus the Milky Way will have: $r_{\rm s,MW}=10.0$~kpc, 
LMC will have $r_{\rm s,LMC}=0.3\cdot r_{\rm s,MW}=3.0$~kpc, 
M31 will have $r_{\rm s,M31}=1.5\cdot r_{\rm s,MW}=15.0$~kpc, 
and M33 will have $r_{\rm s,M33}=0.4\cdot r_{\rm s,MW}=4.0$~kpc. 
\begin{figure}[t,b]
\begin{center}
\begin{tabular}{c@{\hspace{0cm}}c}
(a)\includegraphics[width=6.15cm]{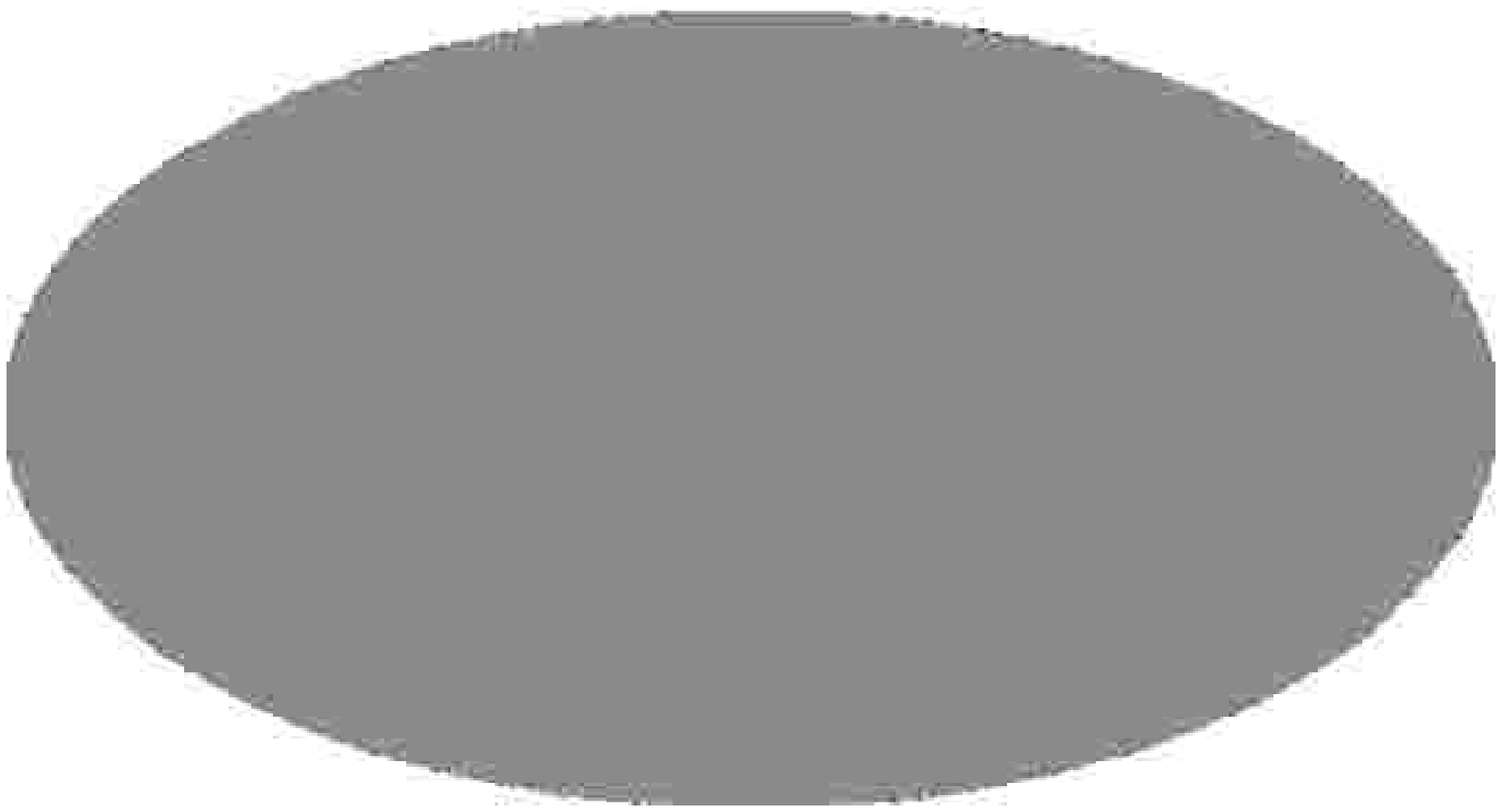}&
(b)\includegraphics[width=6.15cm]{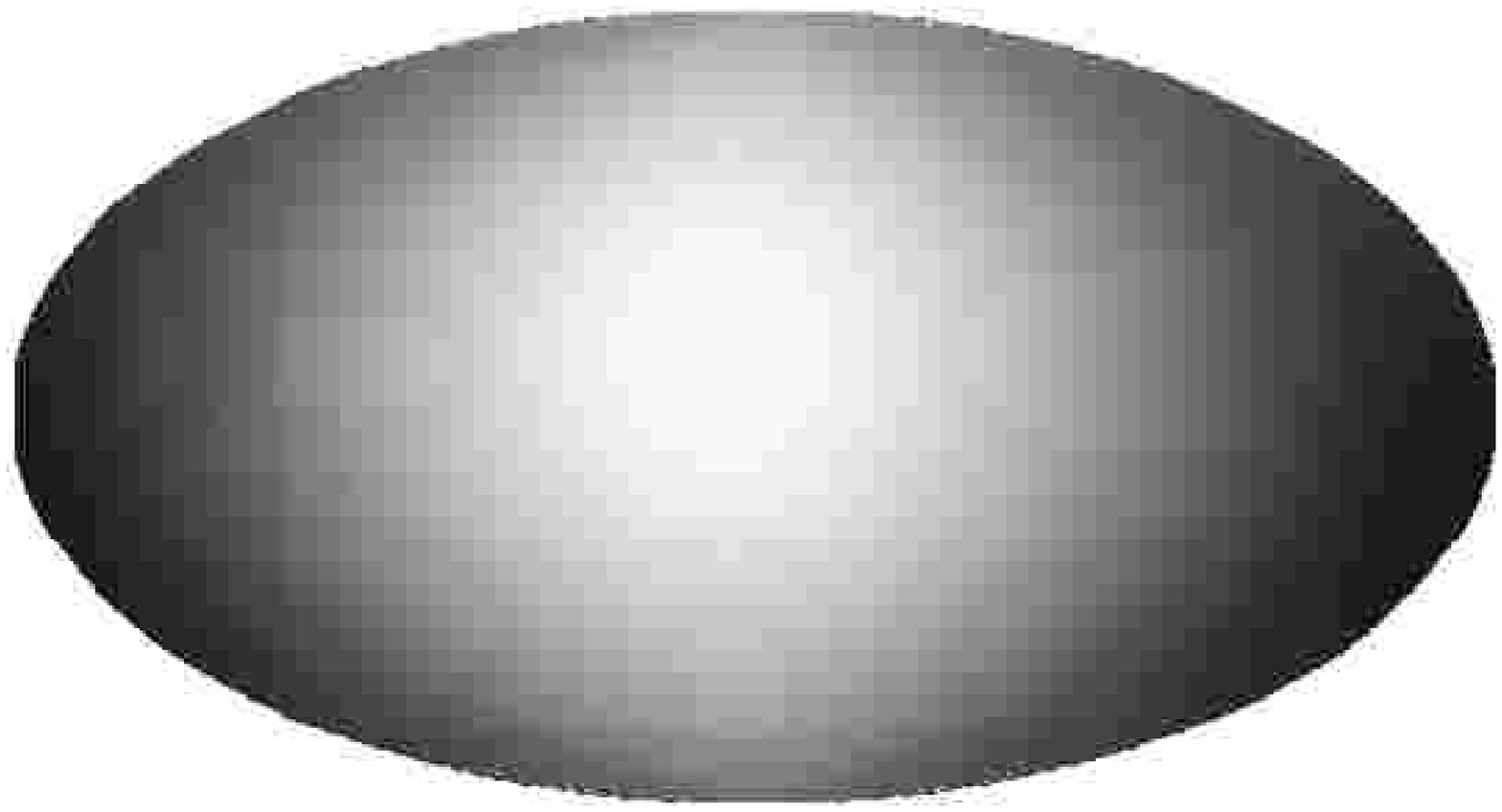}\\
(c)\includegraphics[width=6.15cm]{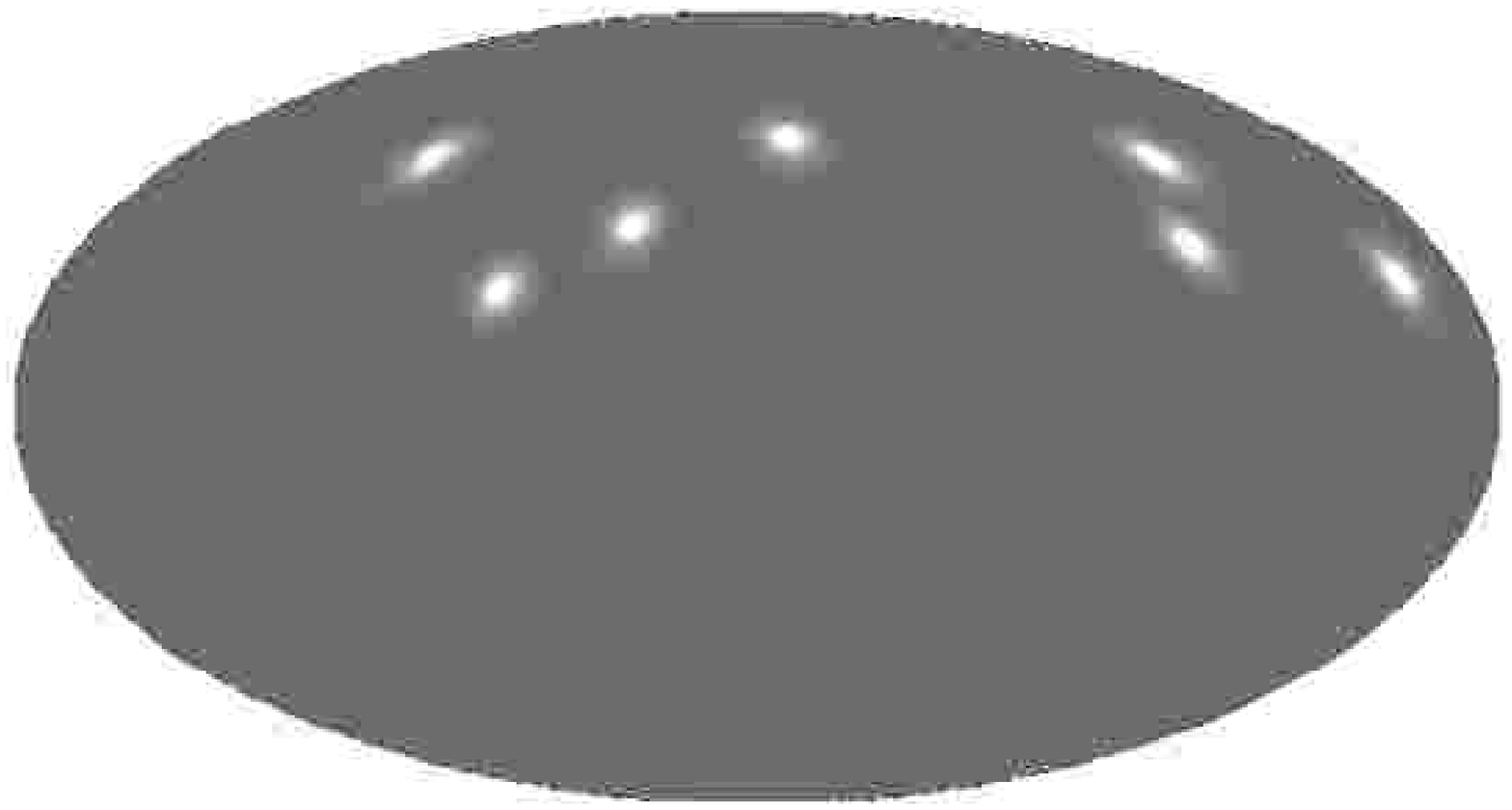}&
(d)\includegraphics[width=6.15cm]{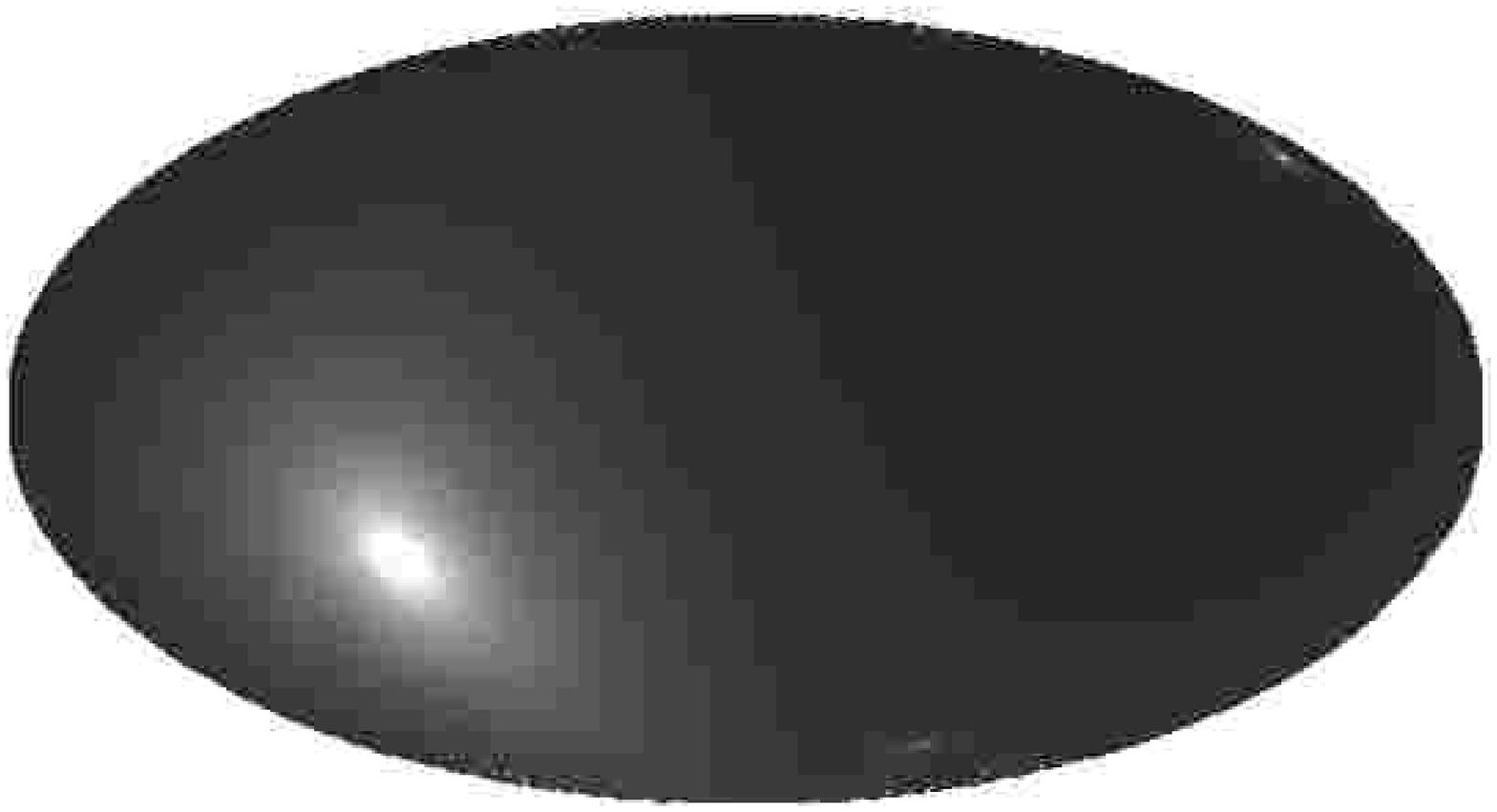}
\end{tabular}
\end{center}
\caption{Density profiles for four different source models---(a):
isotropic model; (b): dipole enhancement model $(\alpha=1.0)$; (c):
seven source model $(F_s=0.28)$; 
(d): dark matter halo model $(r_s=10~{\rm kpc})$. 
All four figures are shown in a Hammer-Aitoff projection of equatorial coordinates (right ascension right
to left).
The highest density in each panel corresponds to the lightest (red)
regions, the lowest density to the darkest (blue) regions.}
\label{figure:source}
\end{figure}

We now calculate the information dimension, $D_{\rm I}$,
 for each of the our four models.  Since these 
are smooth distributions with no statistical fluctuations, we will only use
one, computationally limited value for the number of declinational bands
for each model of $N_{\rm\delta}=1080$ 
(i.e. $\Delta\theta=\frac{1}{6}^\circ$), which implies:
\begin{equation}
P_{\rm i}=
n_{\rm i}\biggr[\sum_{\rm i}n_{\rm i}\biggr]^{-1}.
\label{equation:Pi}
\end{equation}  
Using equations~\ref{equation:Di1}~and~\ref{equation:Pi} 
we can now calculate $D_{\rm I}$ for each of the four models.  The
results are in table~\ref{table:Di2}~column~1.  
Note that three of the four values of $D_{\rm I}$ exceed the 
analytical limit of $2$ for a 2-D surface.

If we consider the analytic limit for the isotropic case, we have:  
\begin{equation}
D_{\rm I}=-4\frac{(N_{\rm\delta})^{\rm 2}}{\pi}\frac{P_{\rm i}\log P_{\rm i}}
{\log N_{\rm\delta}}.
\end{equation}
If we then substitute in equation \ref{equation:Pi} we get:
\begin{equation}
D_{\rm I}=\frac{2\log N_{\rm\delta}+\log4/\pi}{\log N_{\rm\delta}} ;
\lim_{N_{\rm\delta} \to \infty}D_{\rm I}(N_{\rm\delta})=2.
\end{equation}
The reason the we obtain values greater than $2$
is because we are working with a finite number of elements on a surface where
the total area does not equal $(N_{\rm\delta})^2$.
\begin{table}[t,b]
\begin{center}
Intrinsic $D_{\rm I}$ values for the different source models that we 
examined in figures~\ref{figure:source}~and~\ref{figure:source_ex}.  
\end{center}
\begin{tabular}{|c|c|c|} \hline
& {\bf 1} & {\bf 2} \\ \hline
& $D_{\rm I}$ for Source Model & $D_{\rm I}$ for Source Model \\
{\bf SOURCE MODEL} & {\bf without} & {\bf with} \\ 
& Detector Exposure & Detector Exposure \\ \hline\hline
Isotropic & 2.035 & 1.967 \\ \hline
Dipole Enhancement & 2.007 & 1.945 \\ \hline
Seven Source & 2.033 & 1.946 \\ \hline
Dark Matter Halo Model & 1.999 & 1.978 \\ \hline
\end{tabular}
\\
\caption{The estimated values of $D_{\rm I}$ with $N_{\rm\delta}=1800$ 
for the four proposed source models for the entire sky 
independent of any real detector's aperture (1) and for the four source
models superimposed on the estimated aperture of an air-fluorescence 
detector at $40^\circ~N$.  These values are mathematical descriptors of a 
data set that are whose number of significant digits are determined by how
extensively each bin is sampled (in this case four digits).  In the case of
a real data set with a finite number of the observations, the number of 
significant digits is constrained by the fluctuations inherent to the 
sample size}
\label{table:Di2}
\end{table}

\subsection{Exposure-Dependent Source Descriptions}
\begin{figure}[t,b]
\begin{center}
\begin{tabular}{c@{\hspace{0cm}}c}
(a)\includegraphics[width=6.15cm]{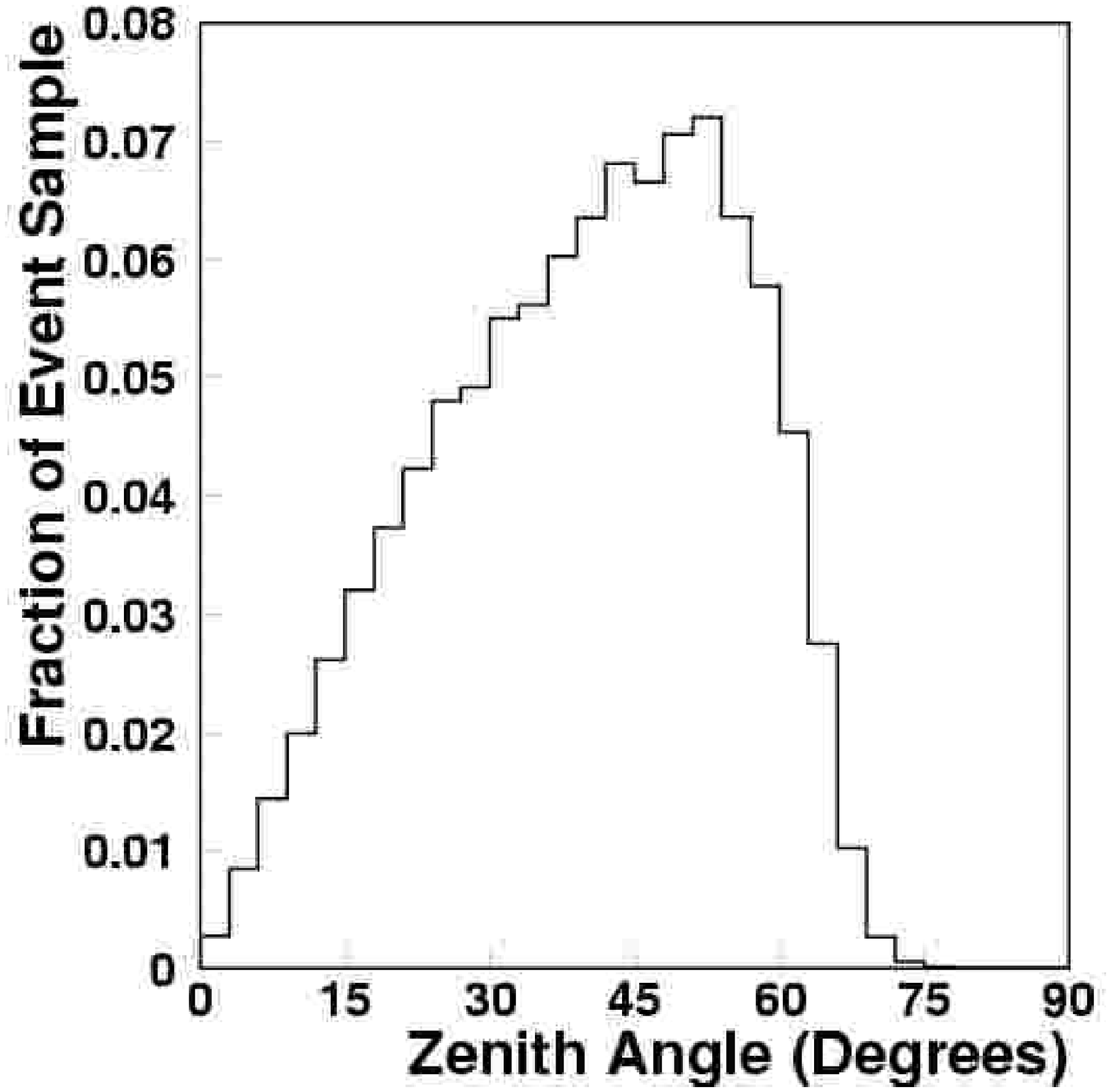}&
(b)\includegraphics[width=6.15cm]{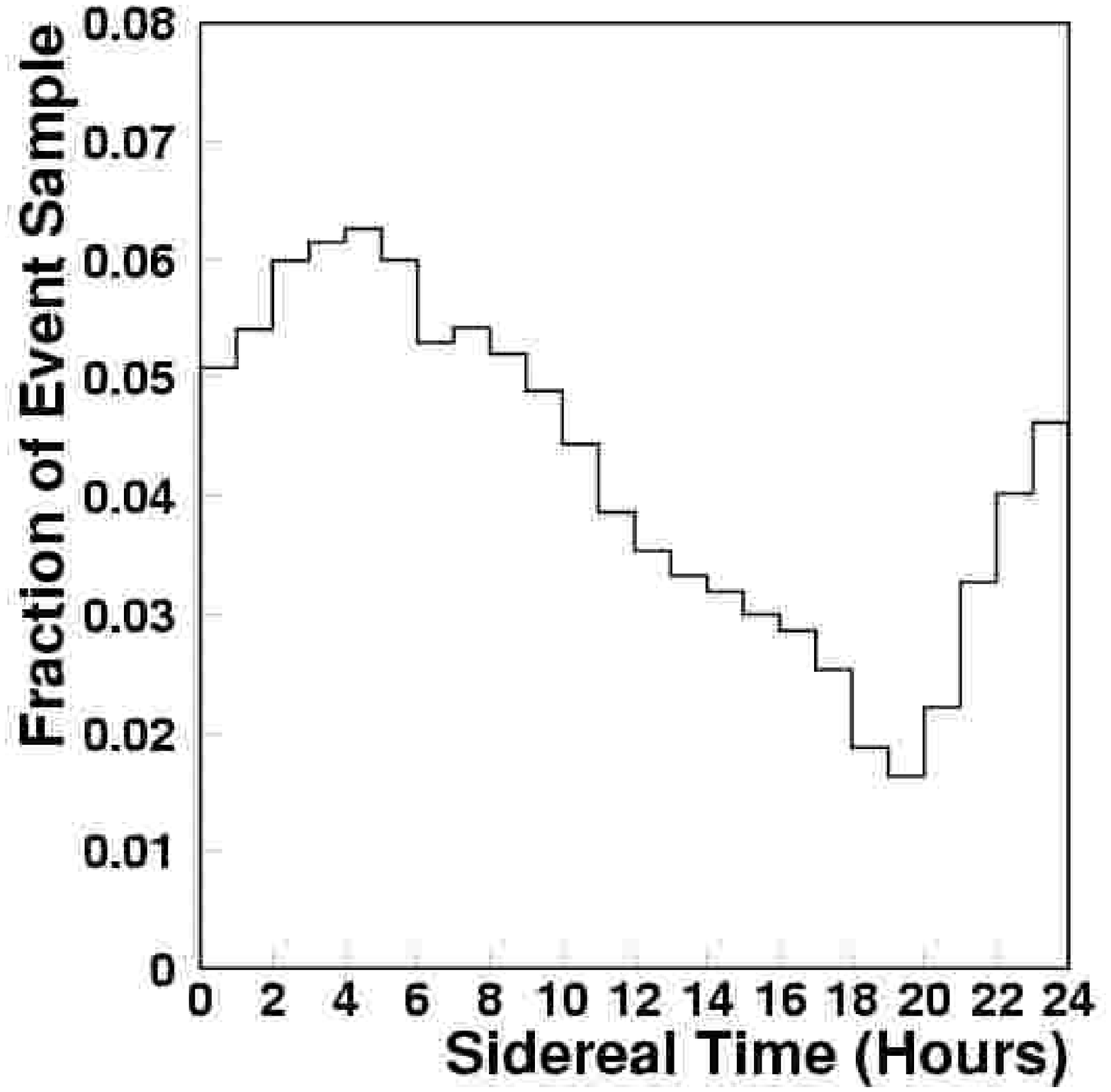}
\end{tabular}
\end{center}
\caption{(a): The distribution of zenith angles 
for a monocular air-fluorescence
detector; (b): the distribution sidereal times for a air-fluorescence
detector located at $40^\circ~N$ in a desert locale (right)}
\label{figure:zensid}       
\end{figure}

For the purpose of this paper, we assume a hypothetical
air-fluorescence detector located at $40^\circ~N$.  This analysis assumes
an isotropic distribution for the azimuthal component of the arrival directions
and a zenith angle distribution that remains constant in time.  This is what
one would expect for a detector with $360^\circ$ coverage with identical 
detector units and stable atmospheric conditions. 
The acceptance of our detector can thus be defined by two distributions: 
zenith angle and sidereal time which are shown in figure~\ref{figure:zensid}.
The zenith angle distribution 
is characterized by $100\%$ acceptance until $\sim50^\circ$, 
at which point it drops off dramatically due to the lack of a well-defined 
profile to assist monocular reconstruction.  The sidereal time distribution
is the combination of the seasonal availability of dark, moonless sky at 
$40^\circ~N$ and
seasonal climatic changes in a desert locale (the rainy season
is assumed to extend from February to May which results in a loss of
$\sim30\%$ of exposure).

\begin{figure}[t,b]
\begin{center}
\begin{tabular}{c@{\hspace{0cm}}c}
(a)\includegraphics[width=6.15cm]{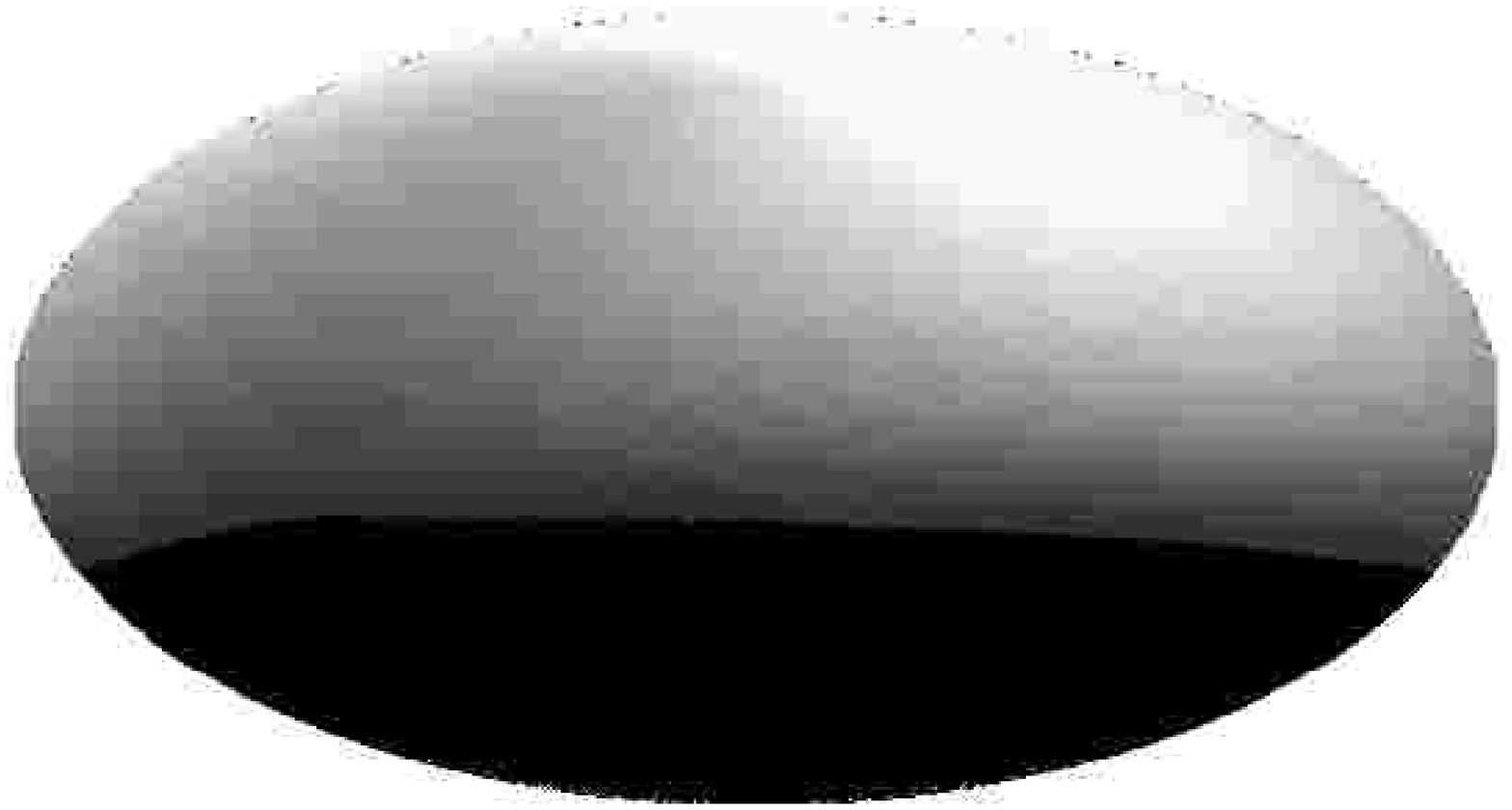}&
(b)\includegraphics[width=6.15cm]{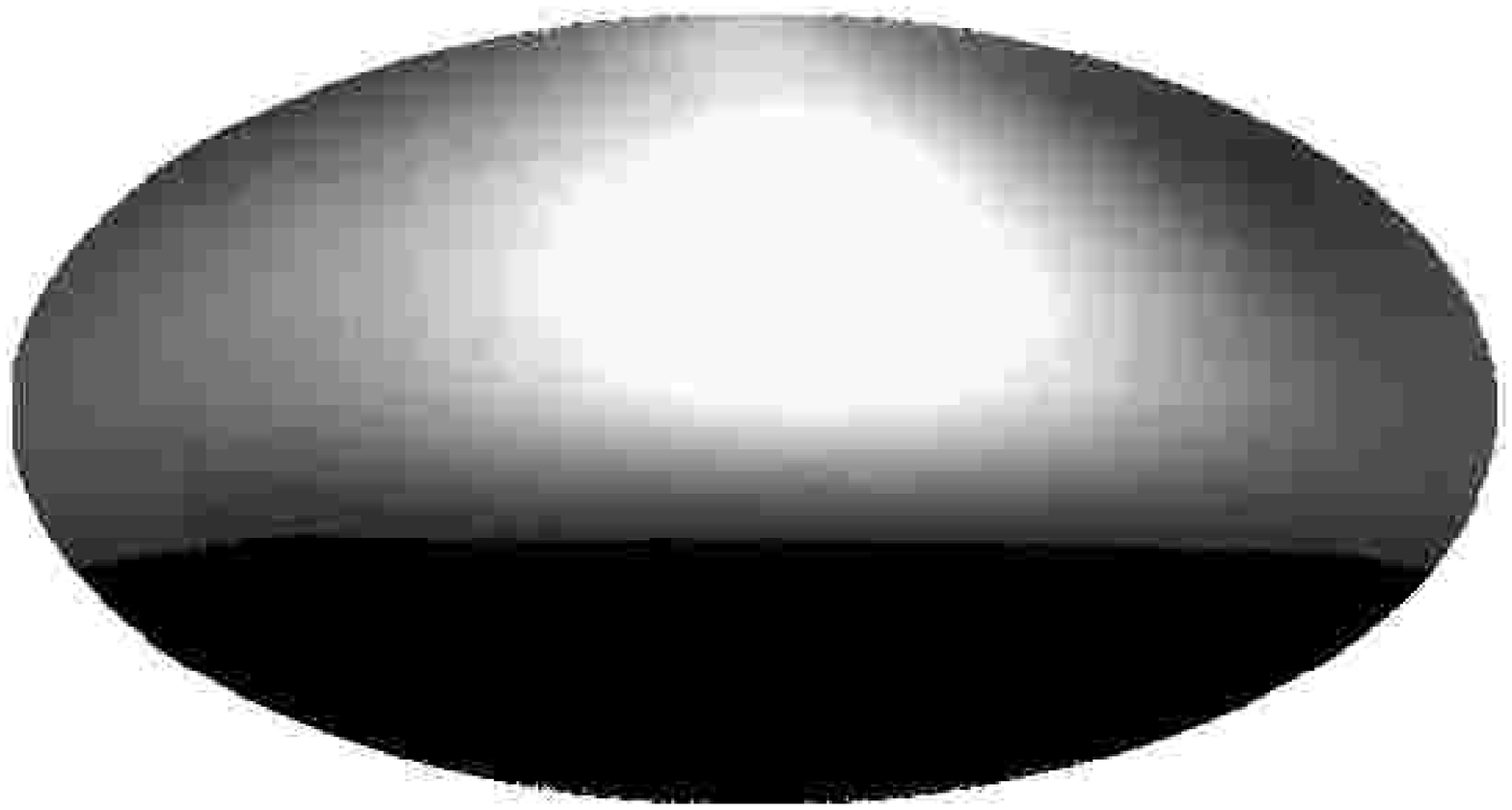}\\
(c)\includegraphics[width=6.15cm]{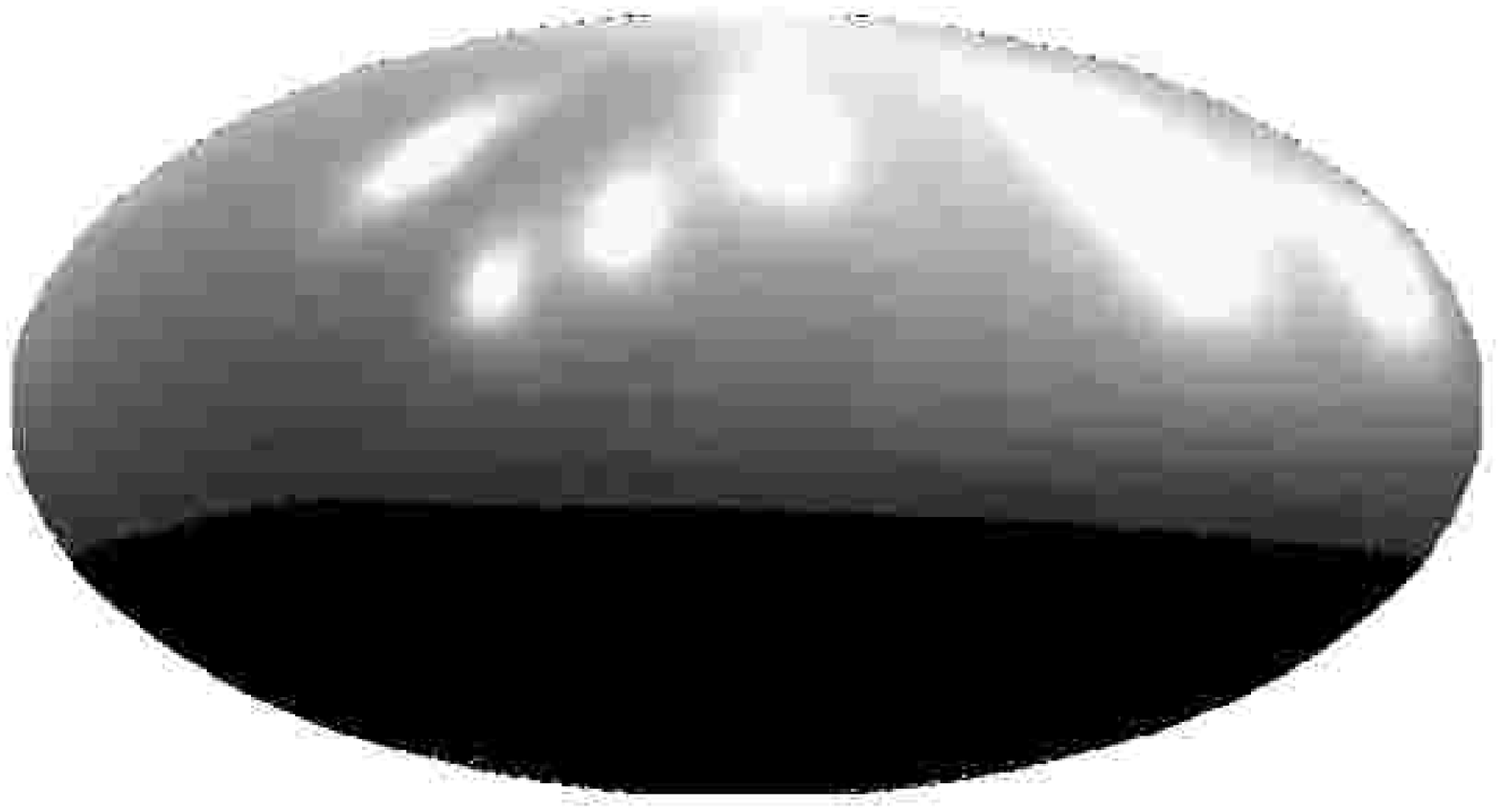}&
(d)\includegraphics[width=6.15cm]{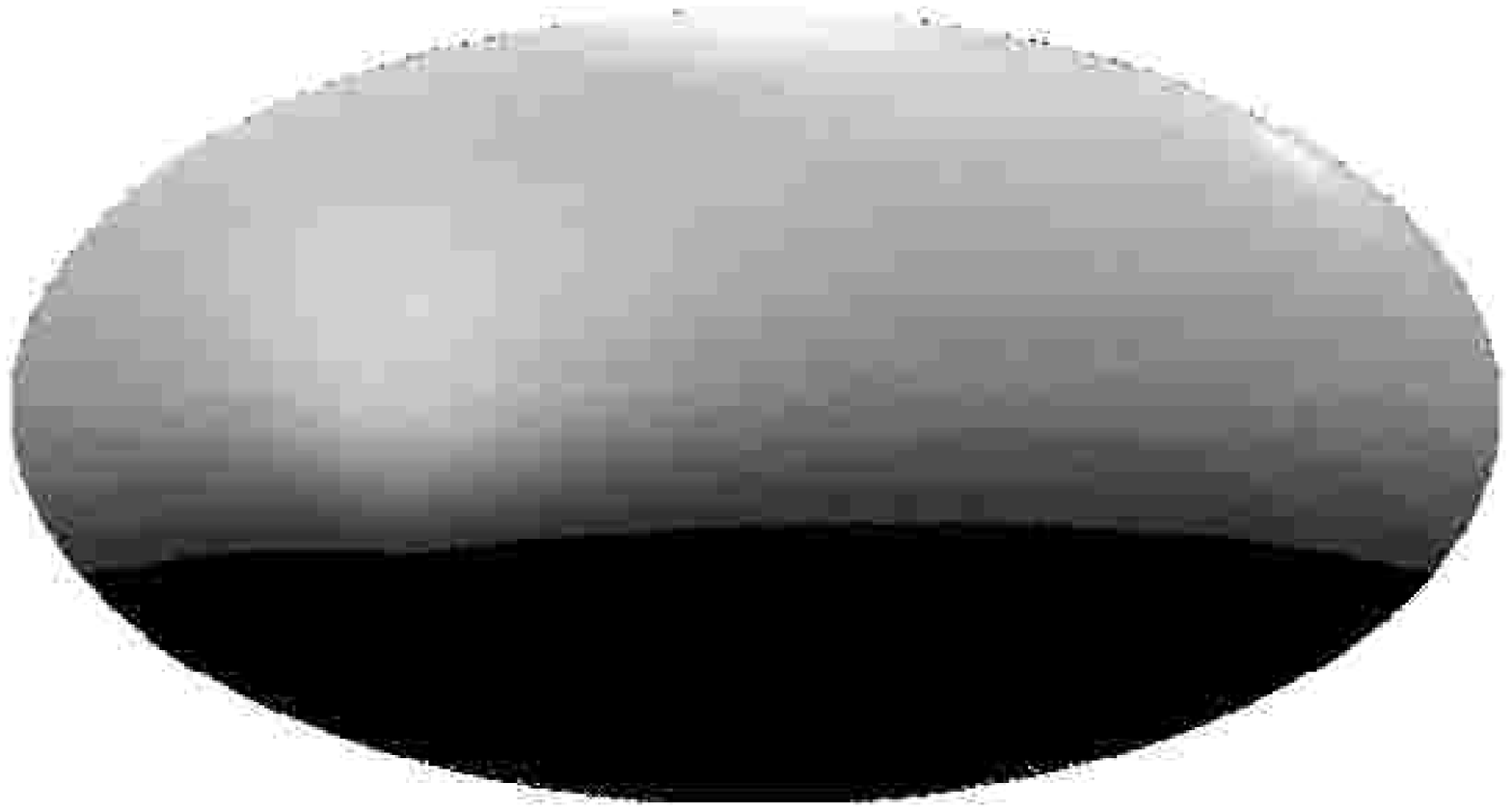}
\end{tabular}
\end{center}
\caption{Effective detector response for the
four different source models---(a): isotropic model; 
(b): dipole enhancement model $(\alpha=1.0)$; (c):
seven source model $(F_s=0.28)$; 
(d): dark matter halo model $(r_s=10~{\rm kpc})$.
All four figures are shown in a Hammer-Aitoff projection of equatorial coordinates (right ascension right
to left).
The highest density in each panel corresponds to the lightest (red)
regions, the lowest density to the darkest (blue) regions.}
\label{figure:source_ex}
\end{figure}

By defining acceptance this way, we can calculate the exposure of the
detector.  By also considering the finite, asymmetric
angular resolution, we can then 
superimpose the detector exposure upon the various source models that we 
previously examined (figure~\ref{figure:source} and 
table~\ref{table:Di2}~column~1). We then obtain an 
effective detector response 
for the air fluorescence detector for each of our
four source models. The results are shown in figure~\ref{figure:source_ex}.
It should be emphasized that effective detector effect is due to the 
{\it combined}
effect of asymmetric sky coverage and angular resolution smearing.  
A remarkable
consequence of this combination is the possibility that point sources can
take on an apparently asymmetric shapes due to preferential orientations of 
the plane of reconstruction for events arriving from a specific location in 
the physical sky.

We can now determine the value of $D_{\rm I}$ using the same method as before.
The results are shown in table~\ref{table:Di2}~column~2.  
It is interesting to note
that the dark matter halo source model now has a larger value for 
$D_{\rm I}$ than the isotropic source model.  By looking at 
figure~\ref{figure:source_ex}, one can verify that the superposition
of the detector exposure and source models actually yields a more uniform
apparent distribution for the dark matter 
halo source model than it does for the 
isotropic source model.  

\section{Calculating {\boldmath$D_{\rm I}$} for Finite Event Samples}

So far, we have only considered calculating $D_{\rm I}$ for smooth 
distributions.  From an experimental standpoint, it is very difficult to 
collect enough data to obtain a smooth distribution.  This is especially
true for UHECRs.  In order to make a measurement of $D_{\rm I}$, 
we must first determine what value(s)
we should assign to $N_{\rm\delta}$.  A reasonable approach is to 
assign a scale length to our sample. Choosing
$\Delta\theta=0.5^\circ$, which approximately reflects the lowest value that 
can be obtained from $\sigma_{\rm2}$ in equation~\ref{equation:sigma2},  
yields $N_{\rm\delta}=360$.  

\begin{figure}[t,b]
\begin{center}
\begin{tabular}{c@{\hspace{0cm}}c}
(a)\includegraphics[width=6.15cm]{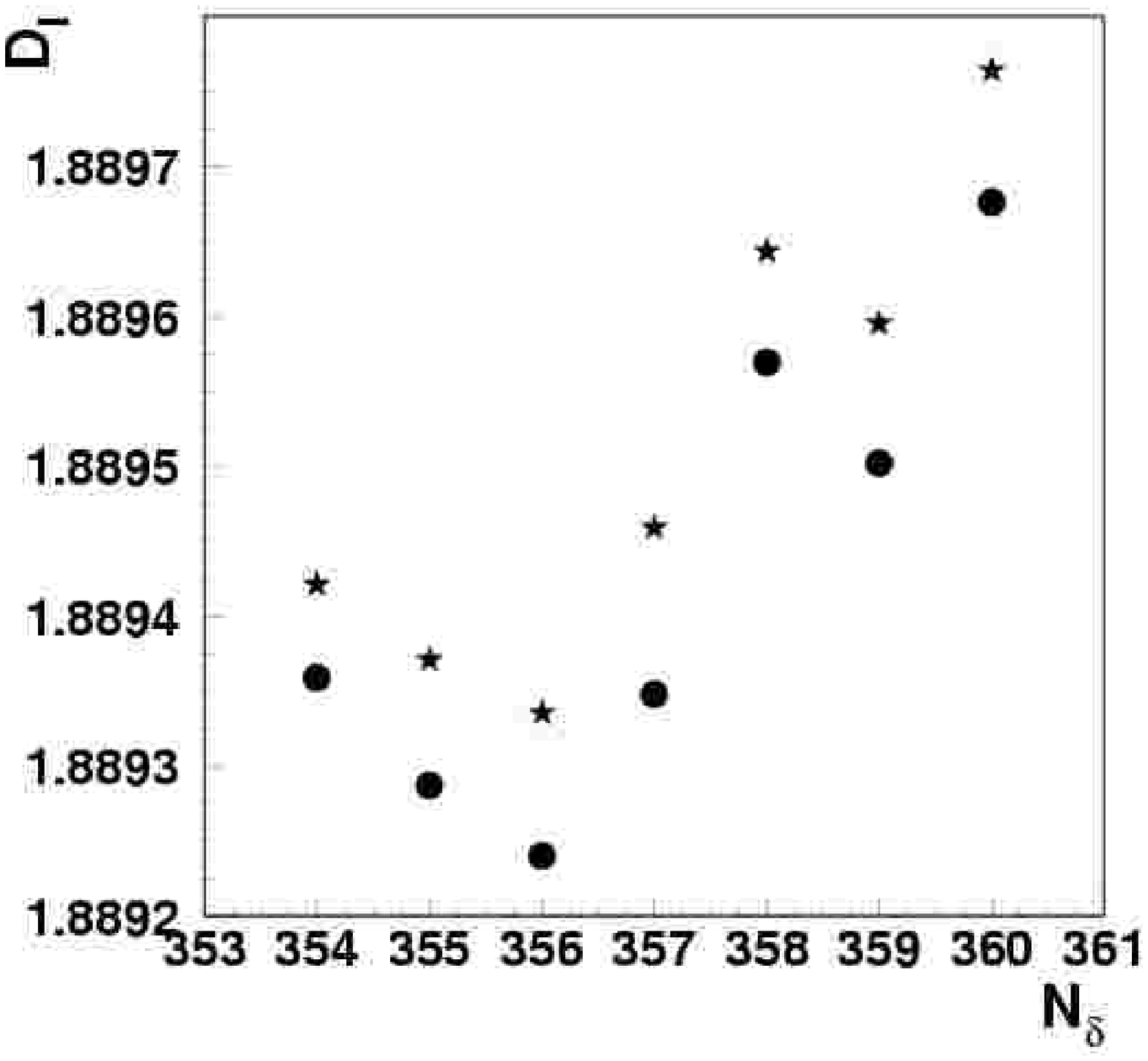}&
(b)\includegraphics[width=6.15cm]{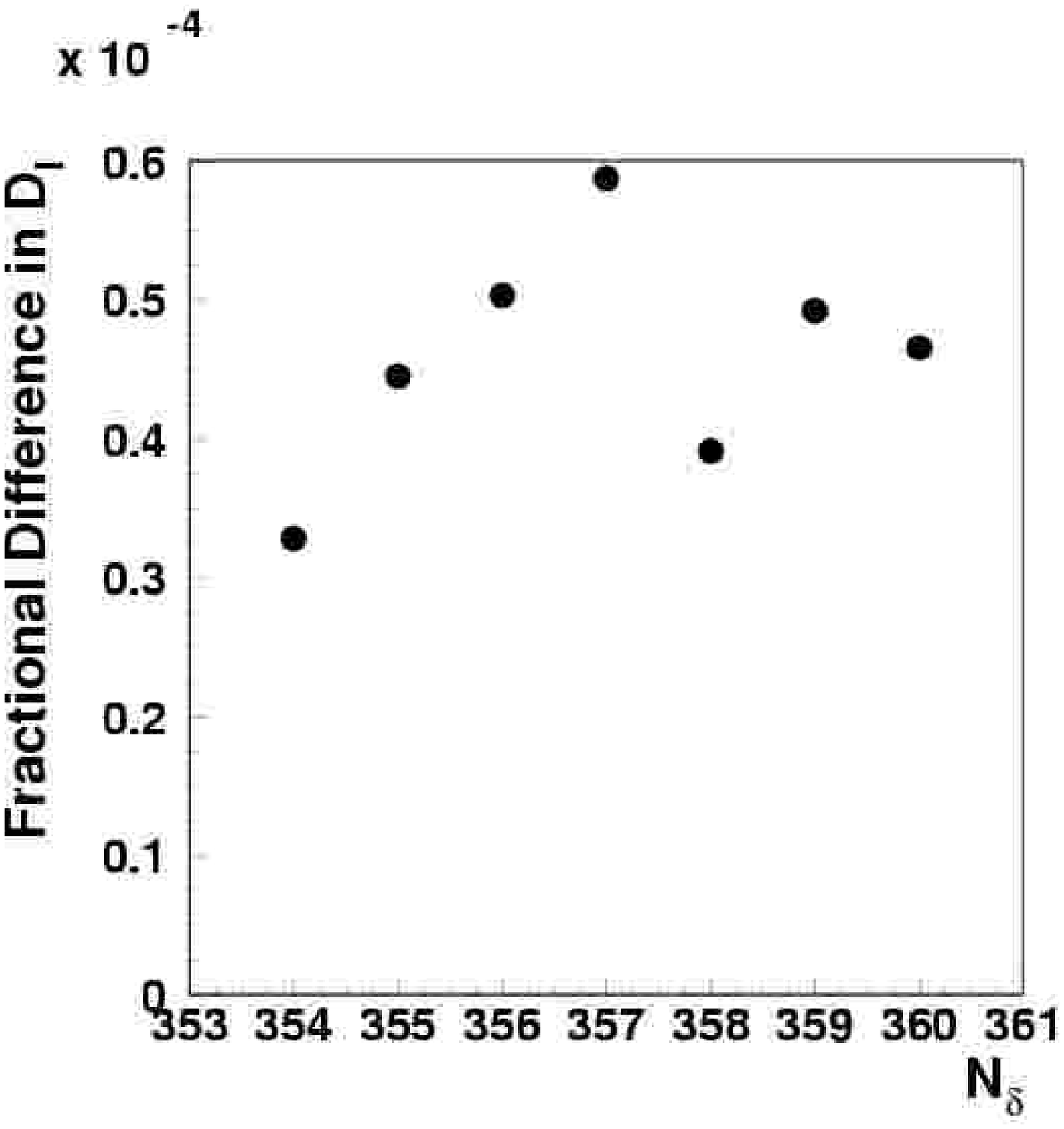}
\end{tabular}
\end{center}
\caption{(a): $D_{\rm I}$ over a range of values of $N_{\rm\delta}$ 
for two similar finite event sets that have similar values for $D_{\rm I}$.
(the dots indicate the values $D_{\rm I}$ for the first set while
the stars indicate the values $D_{\rm I}$ for the second set); 
(b): the fractional difference in $D_{\rm I}$ for the same two sets over the 
range of  $N_{\rm\delta}$ values.}
\label{figure:sp_study}       
\end{figure}

However, it can be beneficial to actually calculate $D_{\rm I}$ for a range
of values around $N_{\rm\delta}$.
In figure~\ref{figure:sp_study}a, 
we display the values 
$D_{\rm I}(N_{\rm\delta})$ over the range $N_{\rm\delta}=[354,360]$ for 
two separate simulated sets where $N_{\rm Shower}=500$.  These sets yield
very similar values for $D_{\rm I}$ over the full range of values for 
$N_{\rm\delta}$. However, if we examine figure~\ref{figure:sp_study}b 
we can see that for the same two finite samples, the 
fractional difference between values of $D_{\rm I}$ can fluctuate 
substantially even over small intervals of $N_{\rm\delta}$.  While these
fluctuations are typically much smaller than the difference in $D_{\rm I}$ 
values between any two sets, we will take $D_{\rm I}$ to be  
$<\!\! D_{\rm I}(N_{\rm\delta})\!\! >$ 
for the interval:  $N_{\rm\delta}=[354,360]$ in order to minimize the 
statistical fluctuations in $D_{\rm I}$ between individual sets.
This range was chosen to optimize our computational ability. 

To account for the 
fact that we no longer have a smooth distribution for the calculation of the 
values of $P_{\rm i}$, we refer back to 
equation~\ref{equation:Pi1}
to see how to calculate $D_{\rm I}$ for a finite set of elements.  This only
requires us to determine a value for $N_{\rm Dist}$.  We can set this value 
based upon what value we wish for $n_{\rm i}$ in combination with
equation~\ref{equation:binnum} 
(i.e. $<\!\! n_{\rm i}\!\! >=\frac{1}{(\Delta n)^{2}}$ 
where $\Delta n$ is the fractional Gaussian fluctuation of a bin
with $n_{\rm i}=<\!\! n_{\rm i}\!\! >$).  Then,
\begin{equation}
N_{\rm Dist}=
\frac{\frac{4}{\pi}(N_{\rm\delta})^{2}<\!\! n_{\rm i}\!\! >}{N_{\rm Shower}}.
\label{equation:ND}
\end{equation}
If we then combine 
equations~\ref{equation:delthet},~\ref{equation:Pi1},~and~\ref{equation:ND}; 
we obtain:
\begin{equation}
P_{\rm i}(N_{\rm\delta})=\frac{n_{\rm i}}{<\!\! n_{\rm i}\!\! >}\frac{\pi^{3}}
{4(N_{\rm\delta})^{4}\Delta\Omega_{\rm\delta}}.
\end{equation}
We can then calculate $D_{\rm I}$ from equation \ref{equation:Di1}:
\begin{equation}
D_{\rm I}=\biggr<\!\! -\frac{1}{\log N_{\rm \delta}}\sum_{i=1}^{N} 
P_{\rm i}(N_{\rm \delta})\log P_{\rm i}(N_{\rm \delta})\! \biggr>,\: 
N_{\rm\delta}=[354,360].
\end{equation}

Thus, If we want $<\!\! n_{\rm i}\!\! >=500$ and $N_{\rm Shower}=500$, we
find that $N_{\rm Dist}\simeq1.65\times10^{5}$.  If $N_{\rm Shower}=2000$, we
have $N_{\rm Dist}\simeq4.1\times10^{4}$. 

\begin{figure}[t,b]
\begin{center}
\begin{tabular}{c@{\hspace{0cm}}c}
(a)\includegraphics[width=6.15cm]{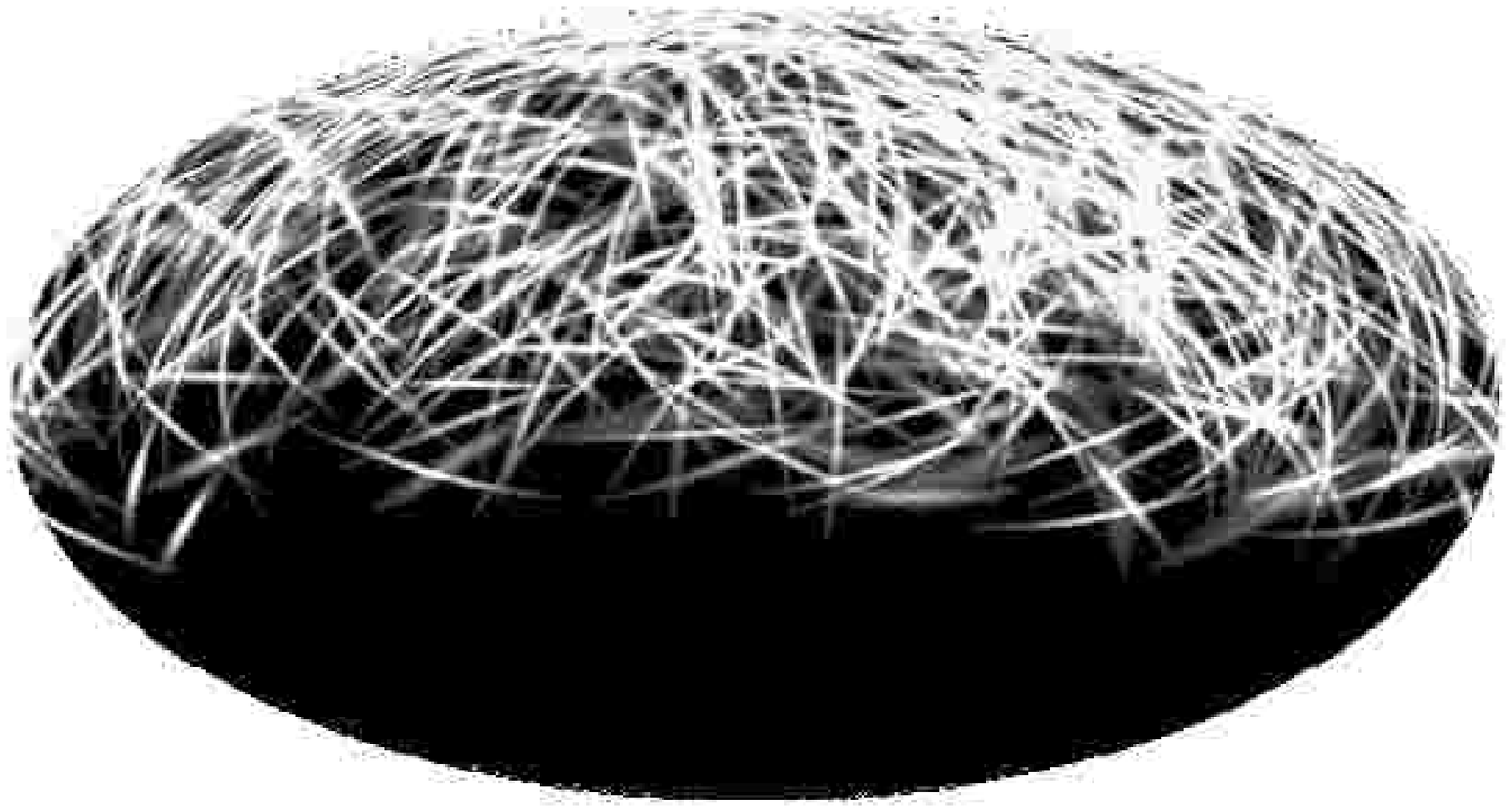}&
(b)\includegraphics[width=6.15cm]{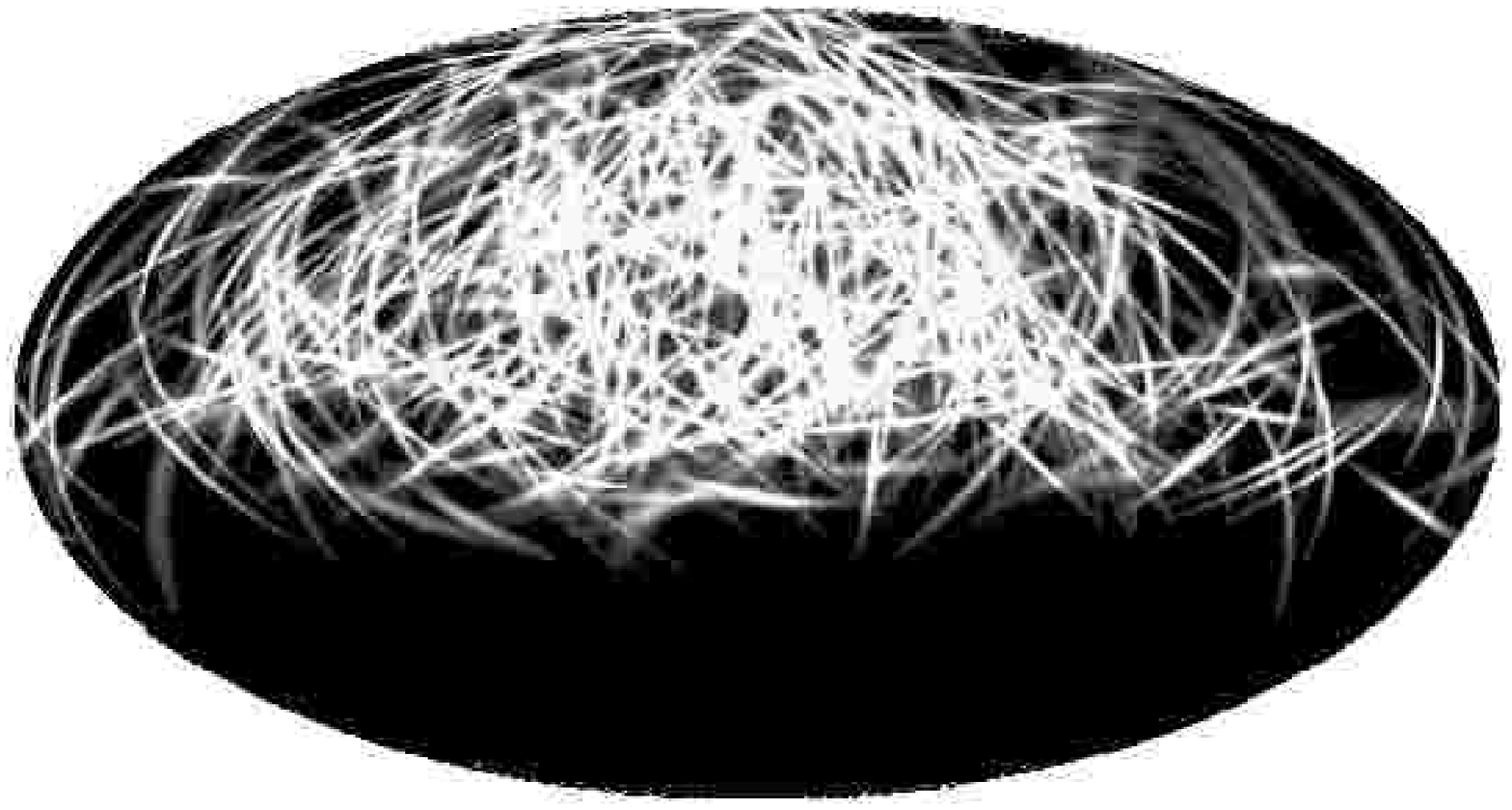}\\
(c)\includegraphics[width=6.15cm]{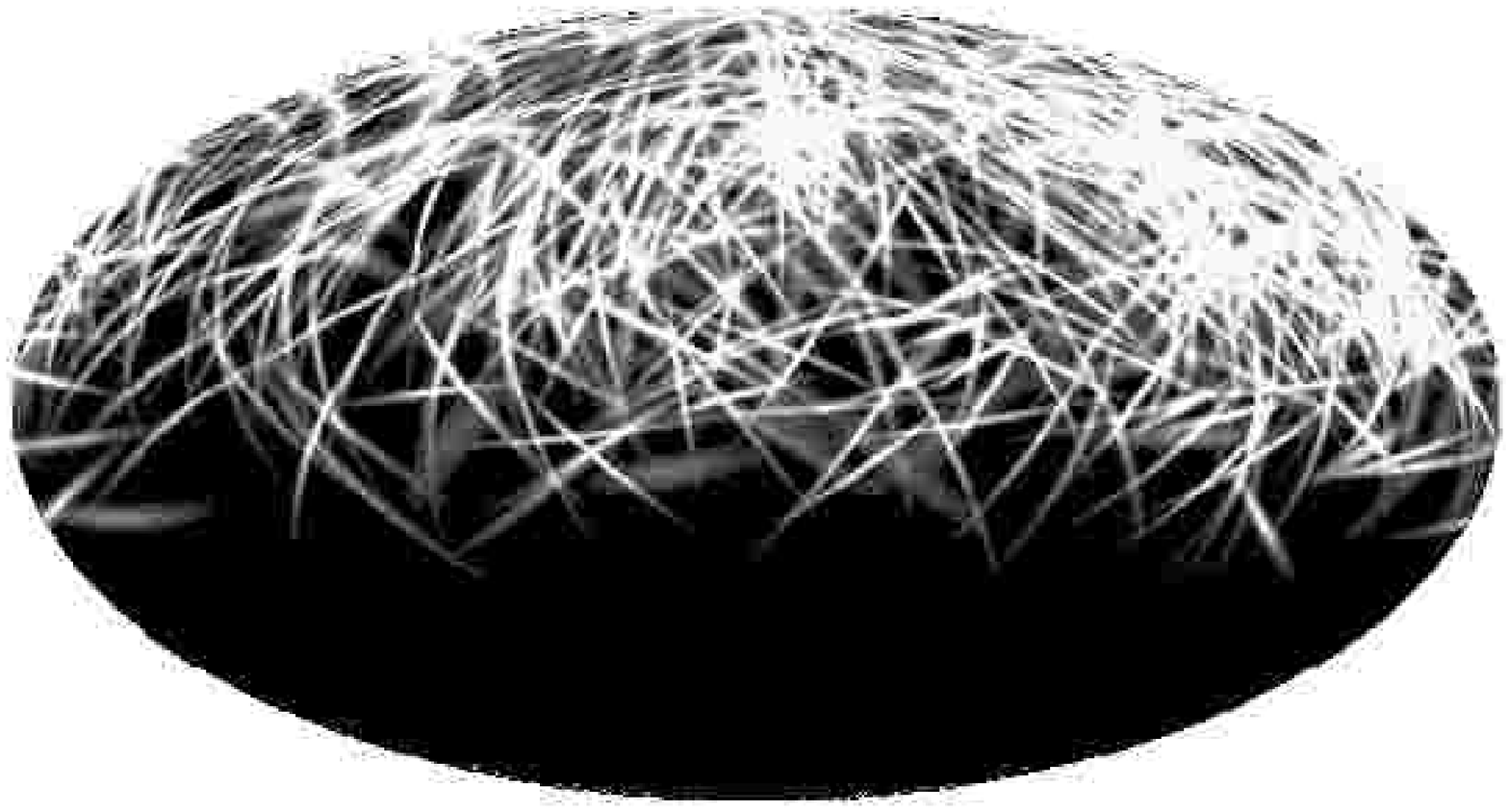}&
(d)\includegraphics[width=6.15cm]{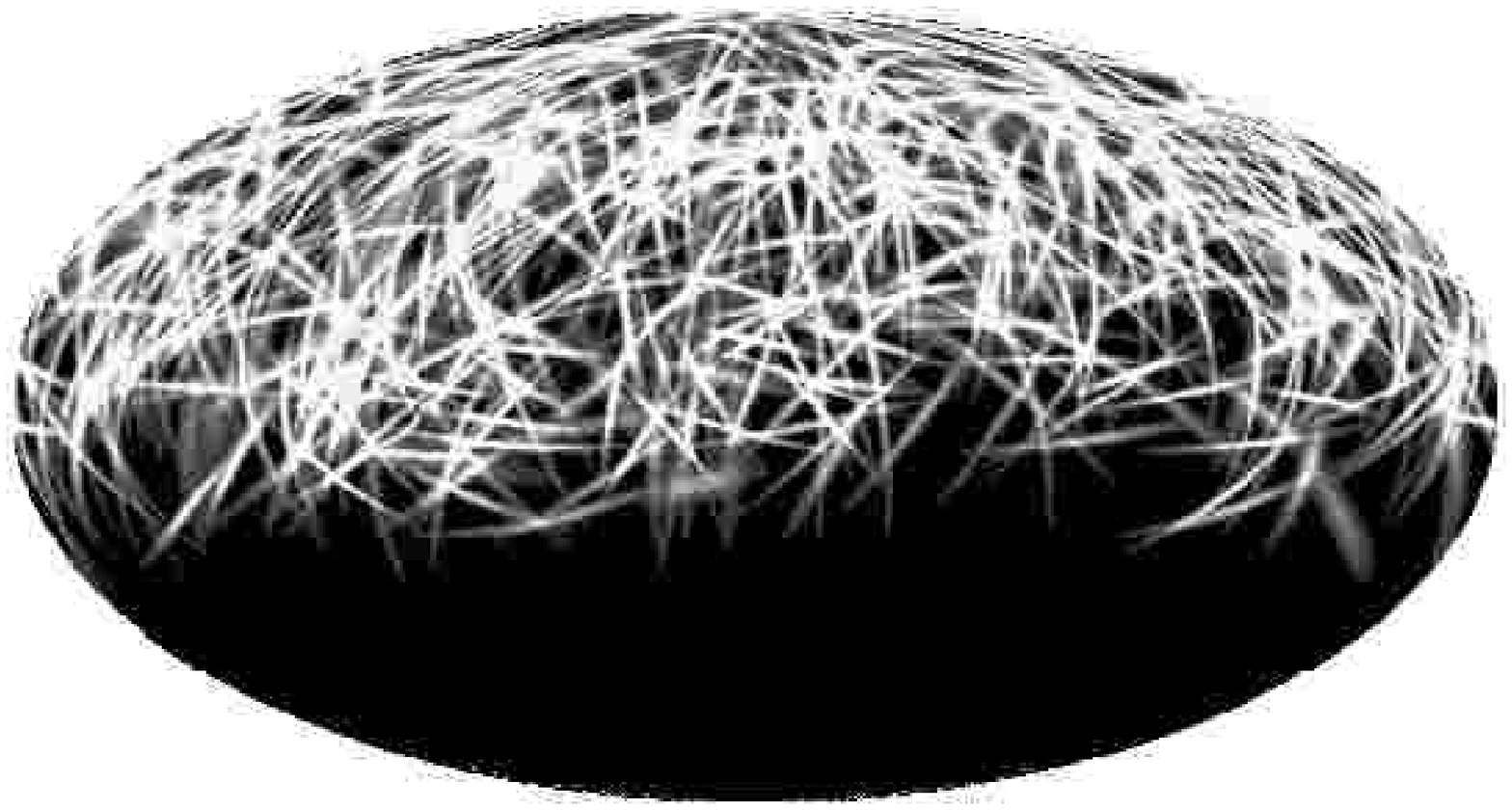}
\end{tabular}
\end{center}
\caption{Simulated 500 event distributions for four different source models 
--- (a): Isotropic model; 
(b): dipole enhancement model $(\alpha=1.0)$; (c):
seven source model $(F_s=0.28)$; 
(d): dark matter halo model $(r_s=10~{\rm kpc})$.
All four figures are shown in a Hammer-Aitoff projection of equatorial coordinates (right ascension right
to left).
The highest density in each panel corresponds to the lightest (red)
regions, the lowest density to the darkest (blue) regions.}
\label{figure:500}
\end{figure}

\begin{figure}[t,b]
\begin{center}
\begin{tabular}{c@{\hspace{0cm}}c}
(a)\includegraphics[width=6.15cm]{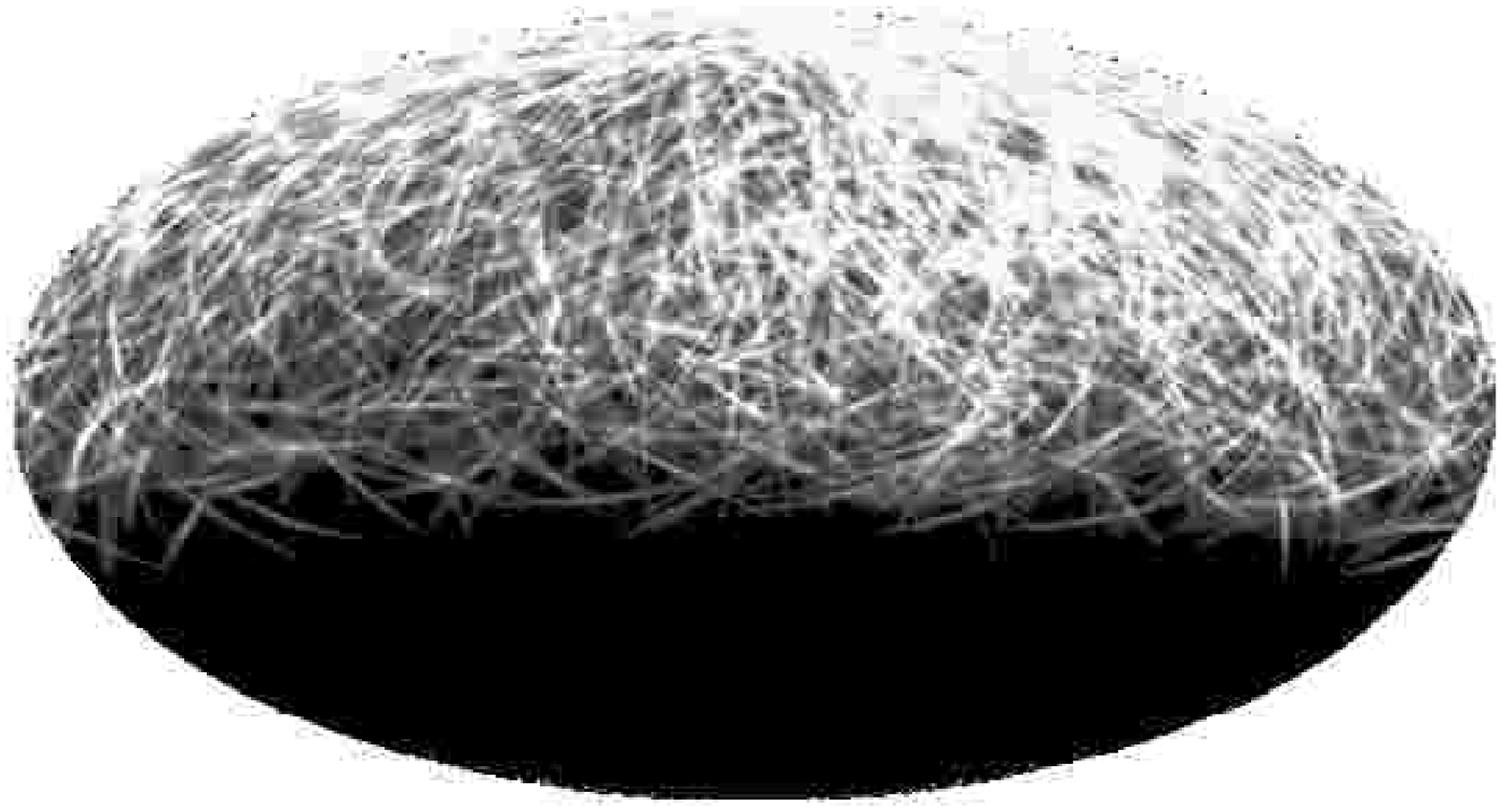}&
(b)\includegraphics[width=6.15cm]{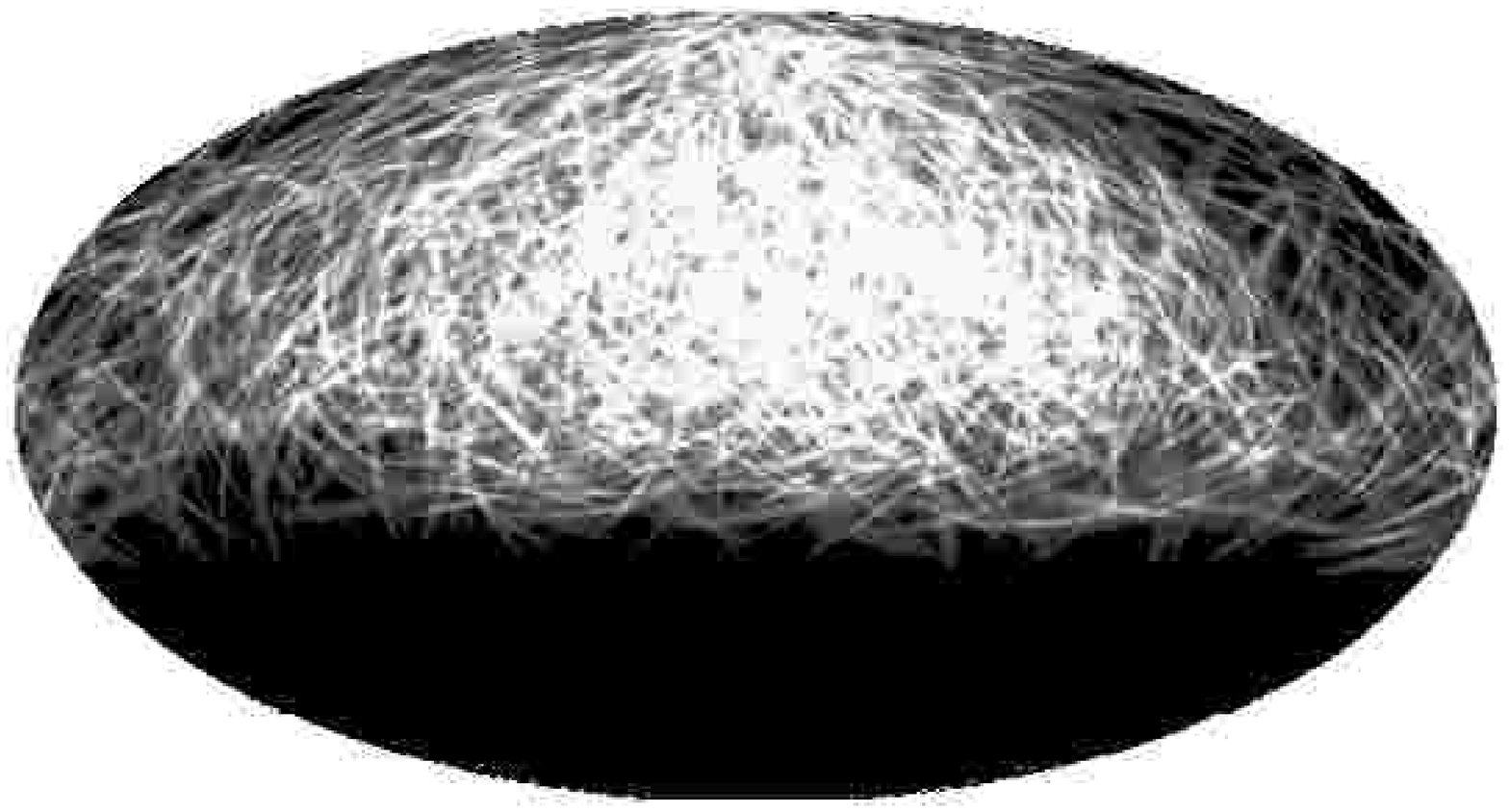}\\
(c)\includegraphics[width=6.15cm]{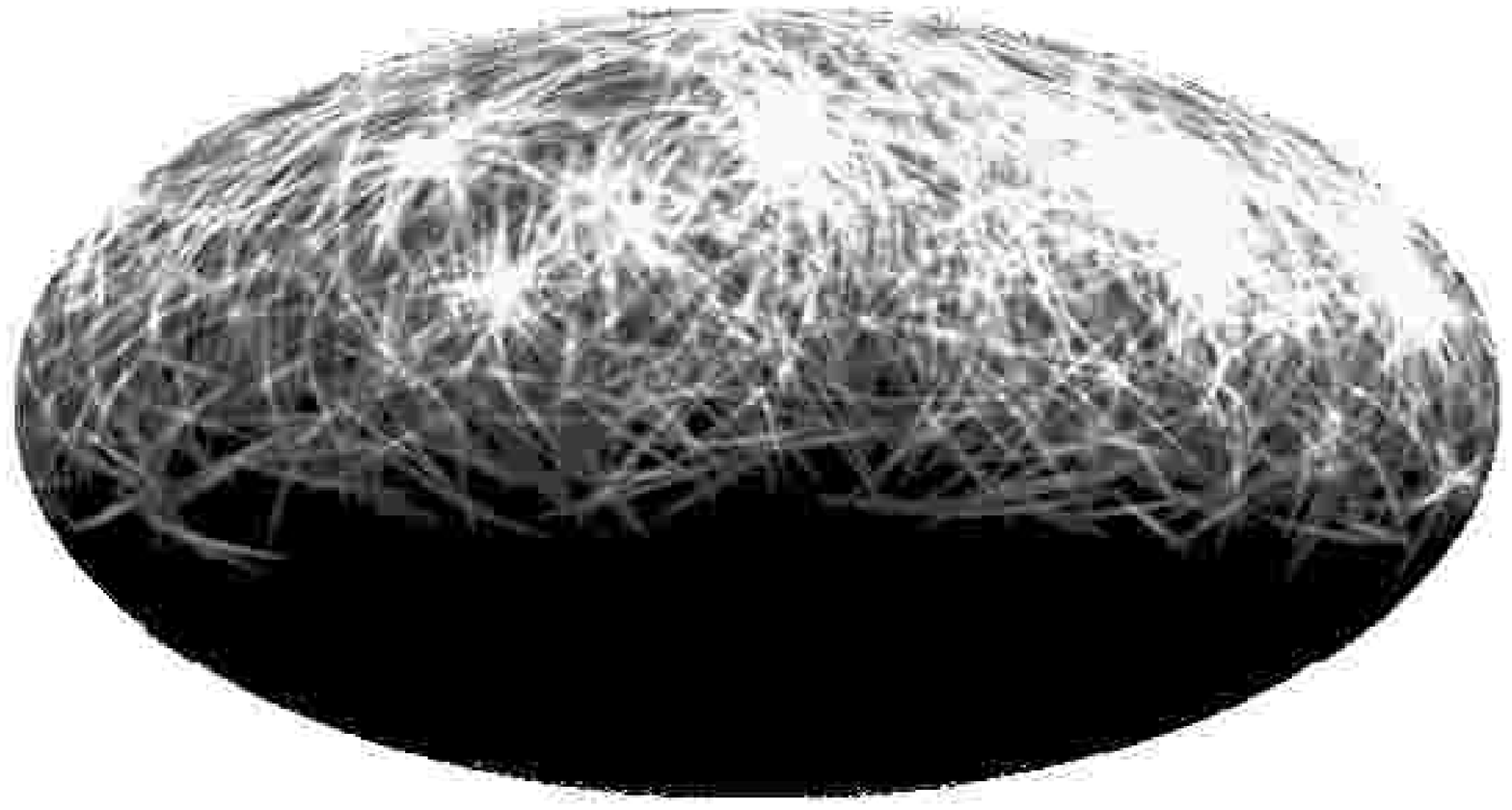}&
(d)\includegraphics[width=6.15cm]{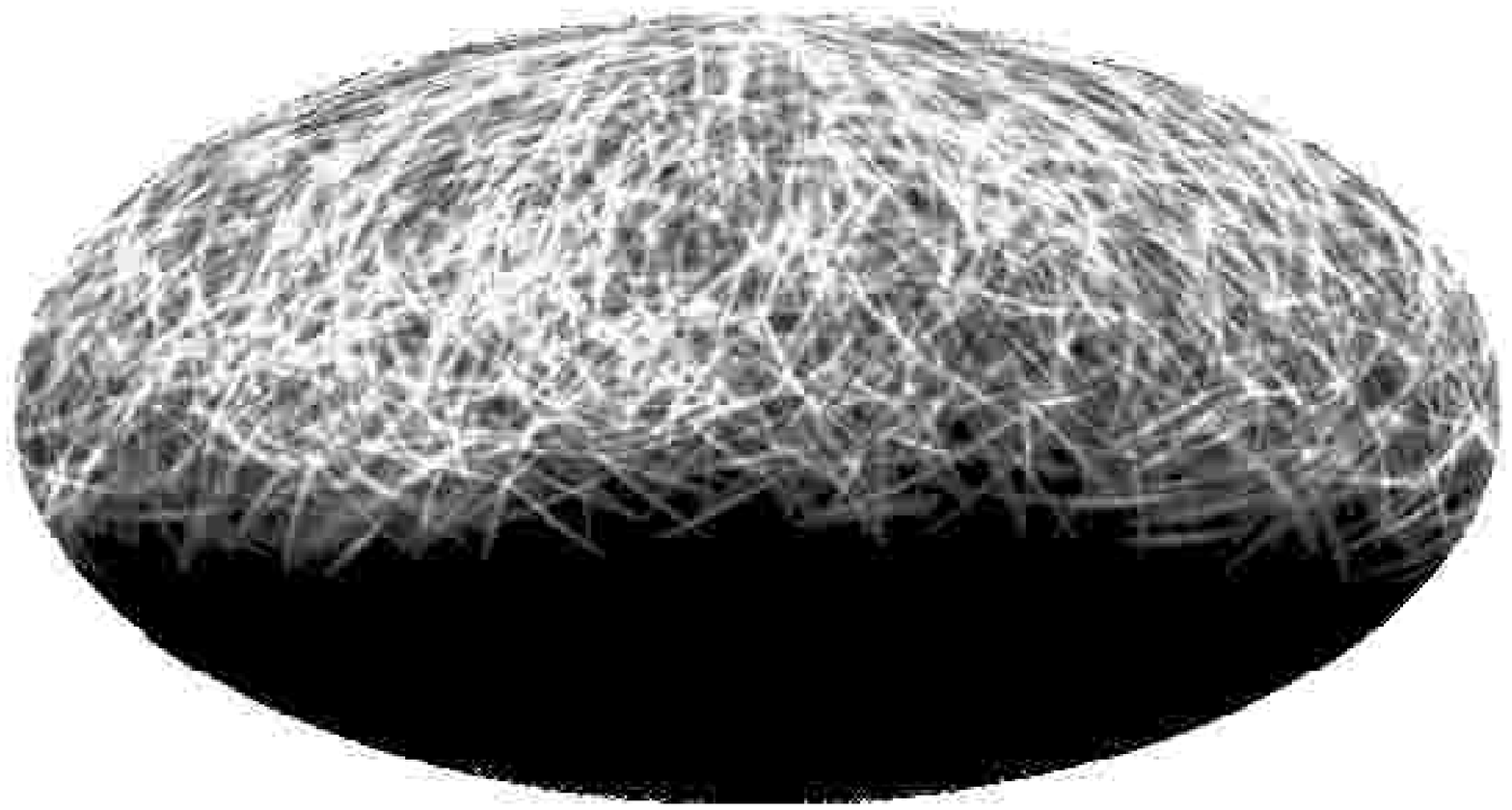}
\end{tabular}
\end{center}
\caption{Simulated 2000 event distributions for four different source models 
--- (a): Isotropic model; (b): dipole enhancement model $(\alpha=1.0)$; (c):
seven source model $(F_s=0.28)$; 
(d): dark matter halo model $(r_s=10~{\rm kpc})$.
All four figures are shown in a Hammer-Aitoff projection of equatorial coordinates (right ascension right
to left).
The highest density in each panel corresponds to the lightest (red)
regions, the lowest density to the darkest (blue) regions.}
\label{figure:2000}
\end{figure}

We now consider two cases: finite event sets with 500 events and finite
event sets with 2000 events.  These sets will have the angular resolution
characteristics described in
 equations~\ref{equation:sigma1}~and~\ref{equation:sigma2}.  
The exposure will be modeled via the zenith angle
and sidereal time distributions shown in figure~\ref{figure:zensid}.
Figure~\ref{figure:500} contains examples of event sets with all four
source models and $N_{\rm Shower}=500$ and figure~\ref{figure:2000} contains 
examples of events sets with all four previously described source models and 
$N_{\rm Shower}=2000$.

\begin{figure}[t,b]
\begin{center}
\begin{tabular}{c@{\hspace{0.0cm}}c}
(a)\includegraphics[width=6.15cm]{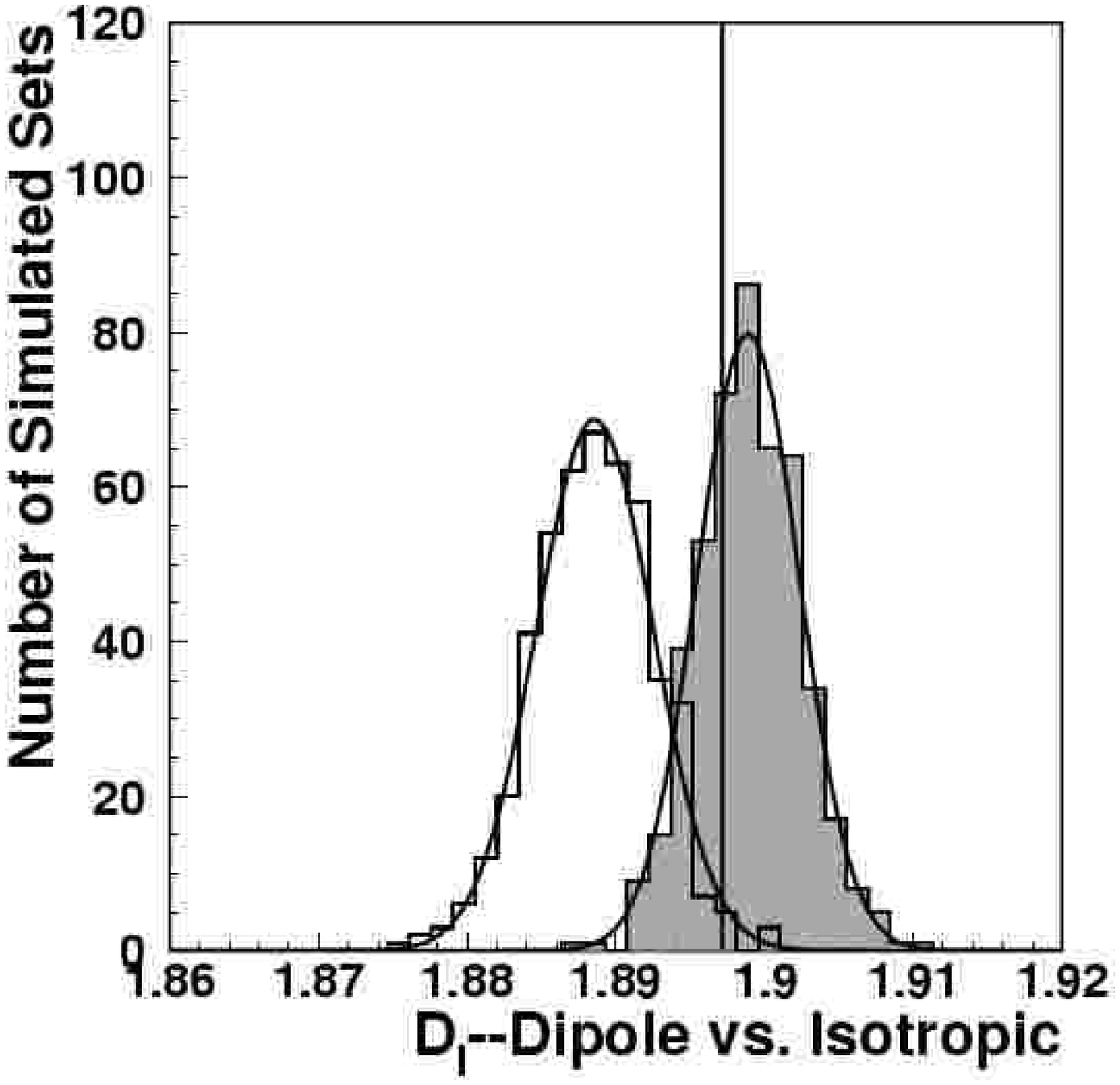}&
(b)\includegraphics[width=6.15cm]{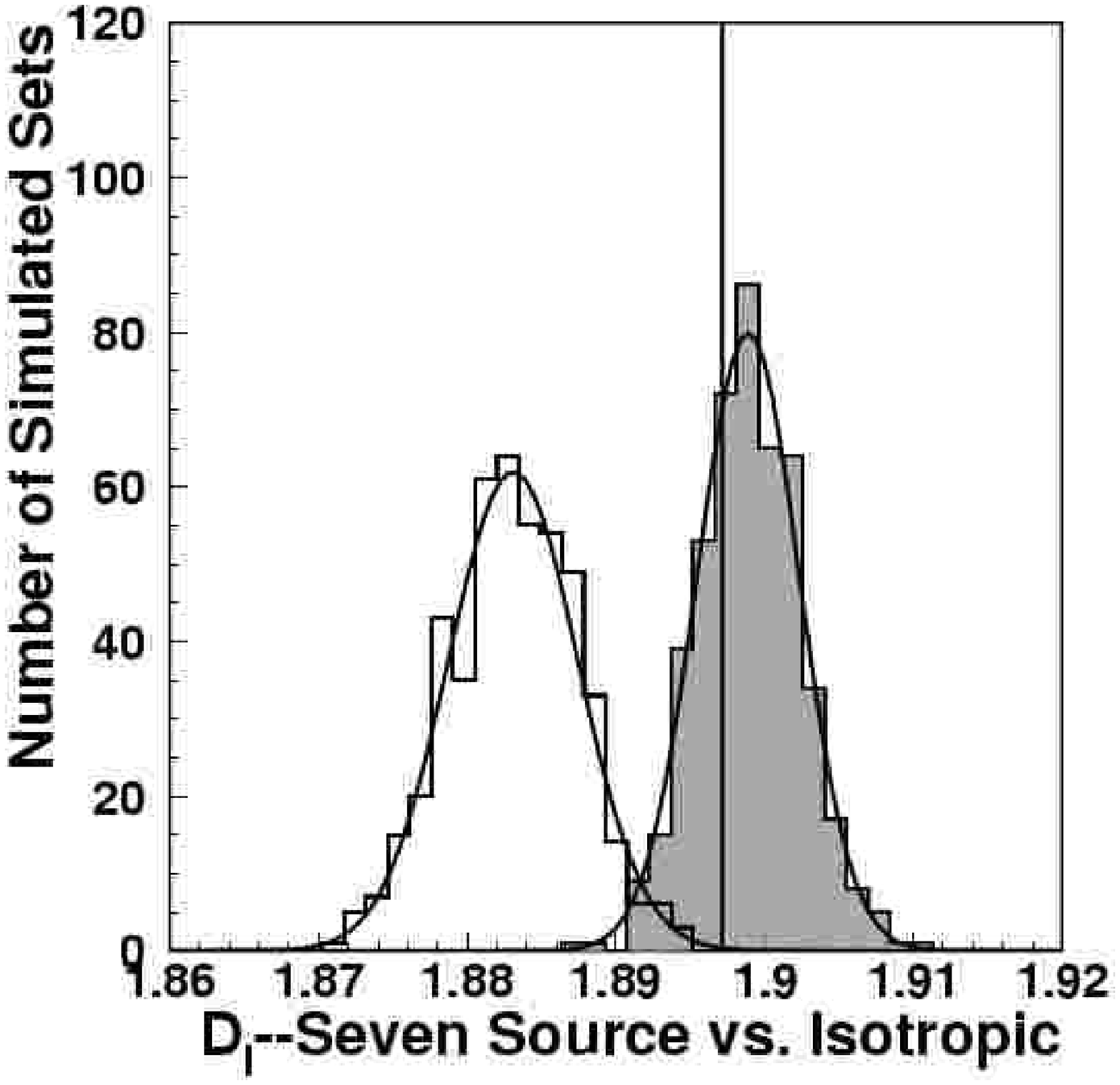}
\end{tabular}
(c)\includegraphics[width=6.15cm]{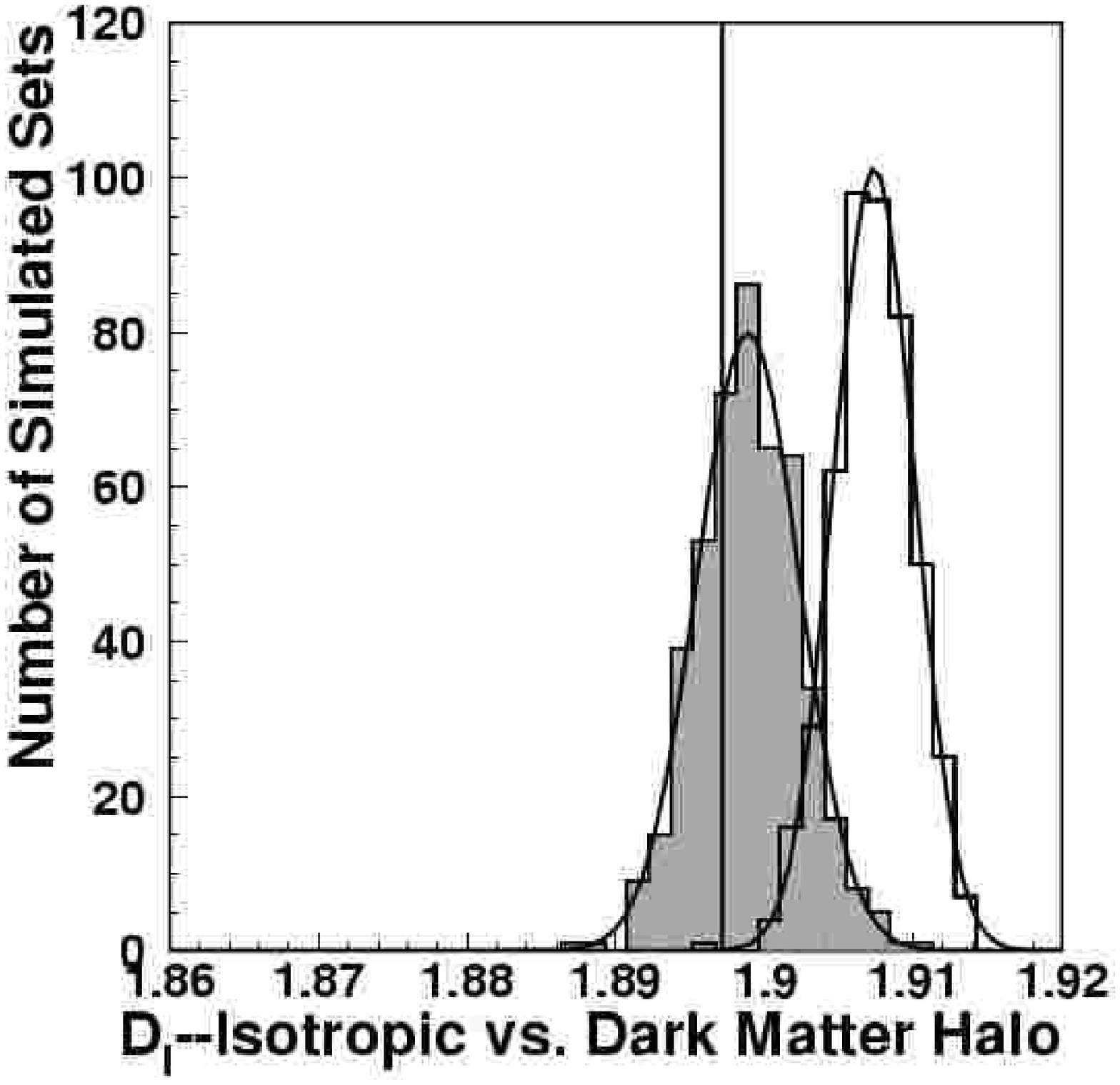}\\
\end{center}
\caption{Comparison of distributions of $D_{\rm I}$ values for $\sim500$ 
sets of 500 events between the isotopic source model (shaded) 
and the other three source models  
--- (a): Dipole enhancement model vs. isotropic model $(\alpha=1.0)$; 
(b): Seven source model vs. isotropic model $(F_s=0.28)$; 
(c): Isotropic model vs. Dark matter halo source model $(r_s=10~{\rm kpc})$.
The vertical line corresponds to the value of $D_{\rm I}$ for the ``real''
500 event sample.  In all cases the distributions of $D_{\rm I}$ values fit 
well to a Gaussian curve.}
\label{figure:500d}
\end{figure}

\begin{figure}[t,b]
\begin{center}
\begin{tabular}{c@{\hspace{0.0cm}}c}
(a)\includegraphics[width=6.15cm]{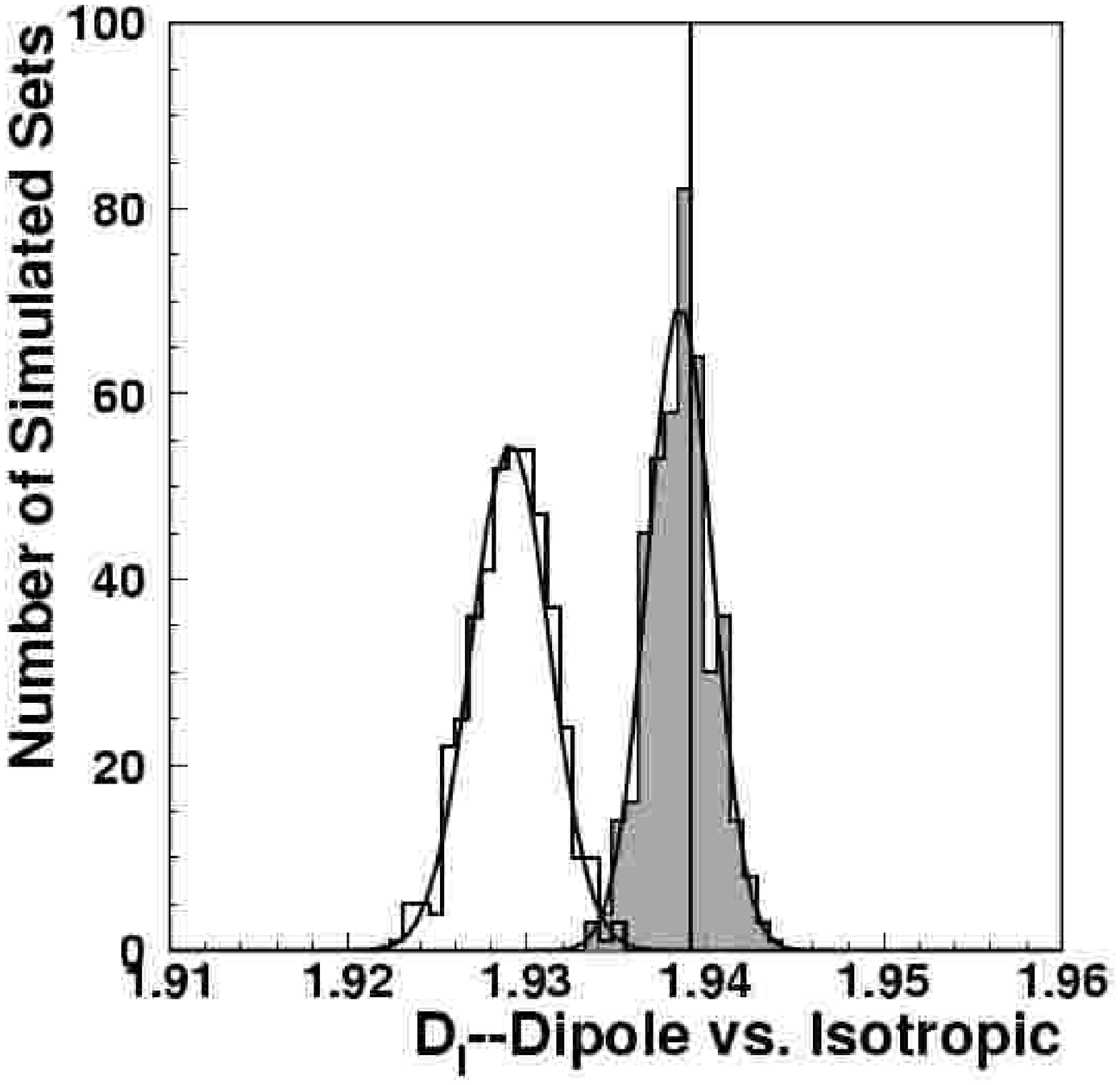}&
(b)\includegraphics[width=6.15cm]{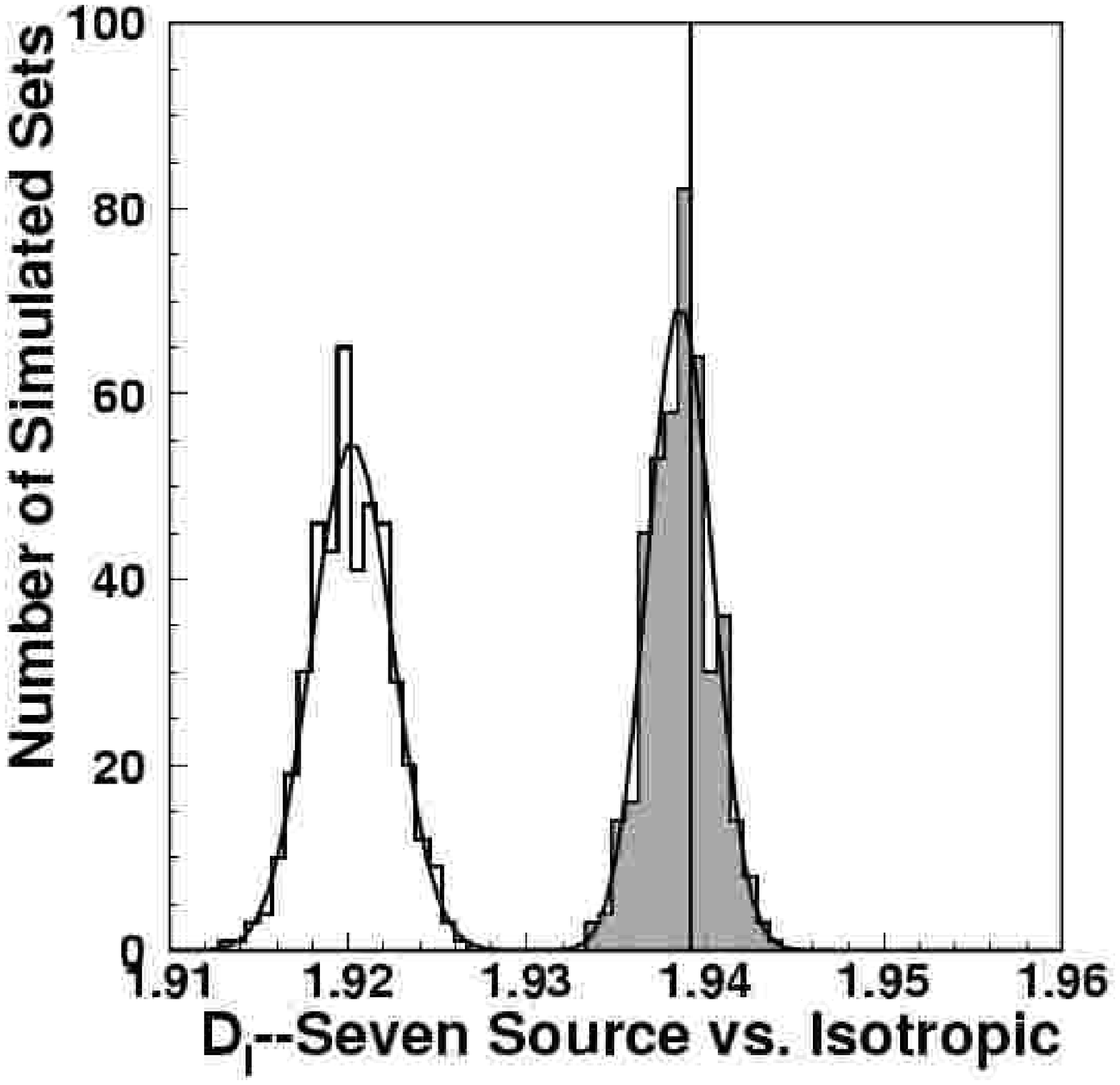}
\end{tabular}
(c)\includegraphics[width=6.15cm]{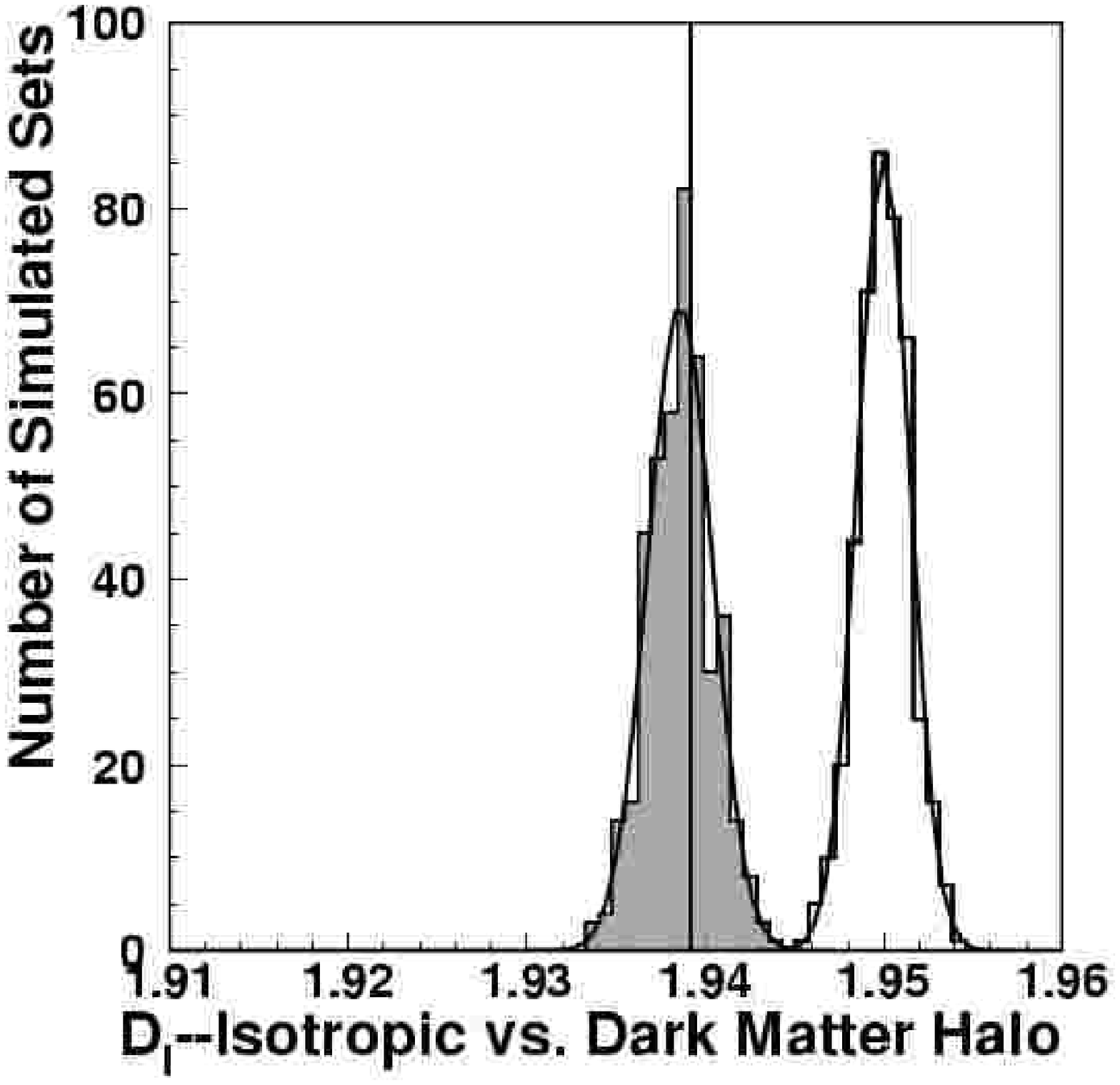}\\
\end{center}
\caption{Comparison of distributions of $D_{\rm I}$ values for $\sim500$ 
sets of 2000 events between the isotopic source model (shaded) 
and the other three source models  
--- (a): Dipole enhancement model vs. Isotropic model $(\alpha=1.0)$; 
(b): Seven source model vs. Isotropic model $(F_s=0.28)$; 
(c): Isotropic model vs. Dark matter halo source model $(r_s=10~{\rm kpc})$.
The vertical line corresponds to the value of $D_{\rm I}$ for the ``real''
2000 event sample.  In all cases the distributions of $D_{\rm I}$ values fit 
well to a Gaussian curve.}
\label{figure:2000d}
\end{figure}

In figures~\ref{figure:500} and \ref{figure:2000} 
one can see the that these distributions of arrival directions have a far
greater degree of statistical fluctuation than the smooth distributions shown
in figure~\ref{figure:source_ex}.
Because of the fluctuations in our simulated event samples,
the value of $D_{\rm I}$ varies significantly 
(see figures~\ref{figure:500d}~and~\ref{figure:2000d})
from one simulated set to the next. In 
figures~\ref{figure:500d}~and~\ref{figure:2000d}, we examine the distribution 
of $D_{\rm I}$ values for $\sim 500$ sets (of 500 and 2000 events 
respectively). We see that the distribution of 500 event samples have both a 
lower mean value and larger width than the 2000 event samples.

\section{Application to Anisotropy Analysis}

We now need to develop a scheme by which we can apply fractal dimensionality
analysis to a real data set.  In the case of a real data set, 
we will be dealing
with only {\it one} value of $D_{\rm I}$.  By itself $D_{\rm I}$ 
is insufficient to characterize the data set;  
$D_{\rm I}$ fluctuates a great deal due to variation in $N_{\rm Shower}$.  
However, a comparison between the value of $D_{\rm I}$ for a real event sample
and a distribution of $D_{\rm I}$ values (with the {\it same}
$N_{\rm Shower}$ and $N_{\rm Dist}$ values as the real data)
 for a series of simulated data sets of a given source model does provide
a viable measurement of anisotropy.

We can demonstrate this by considering a {\it single} simulated event sample 
generated with an isotropic source model.  We will suppose this sample 
to be our 
``real'' data.  We consider the isotropic simulated
event samples shown in figures~\ref{figure:500}a~and~\ref{figure:2000}a.
We will once again stipulate that $<\!\! n_{i}\!\! >=500$
which means that in the case of $N_{Shower}=500$, we have: 
$N_{\rm Dist}\simeq1.6\times10^{5}$ which leads to $D_{\rm I}=1.89715$
and in the case of $N_{Shower}=2000$, we have: 
$N_{\rm Dist}\simeq4\times10^{4}$ which leads to $D_{\rm I}=1.93920$.  

In figures~\ref{figure:500d}~and~\ref{figure:2000d},
we demonstrated that for a fixed scaling parameter
the distribution of $D_{\rm I}$ values for a large number of simulated sets 
fits well to a Gaussian curve.  
By establishing the relationship between the distribution of 
$D_{\rm I}$ values and the scaling parameter in each model, we can establish
a 90\% confidence interval on the scaling parameter for that model.

\subsection{Dipole Enhancement Source Model}

\begin{figure}[t,b]
\begin{center}
\begin{tabular}{c@{\hspace{0.0cm}}c}
(a)\includegraphics[width=6.15cm]{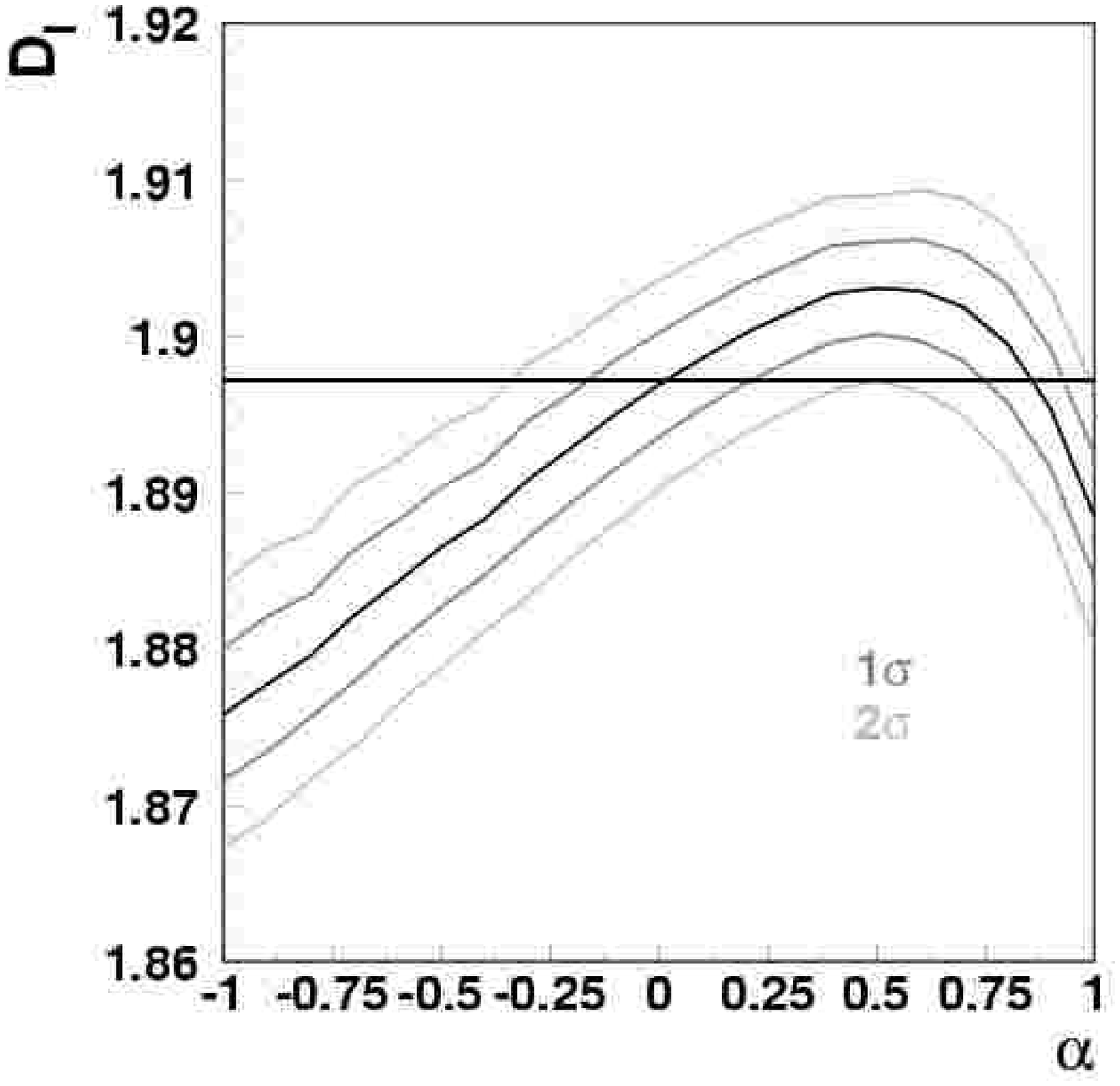}&
(b)\includegraphics[width=6.15cm]{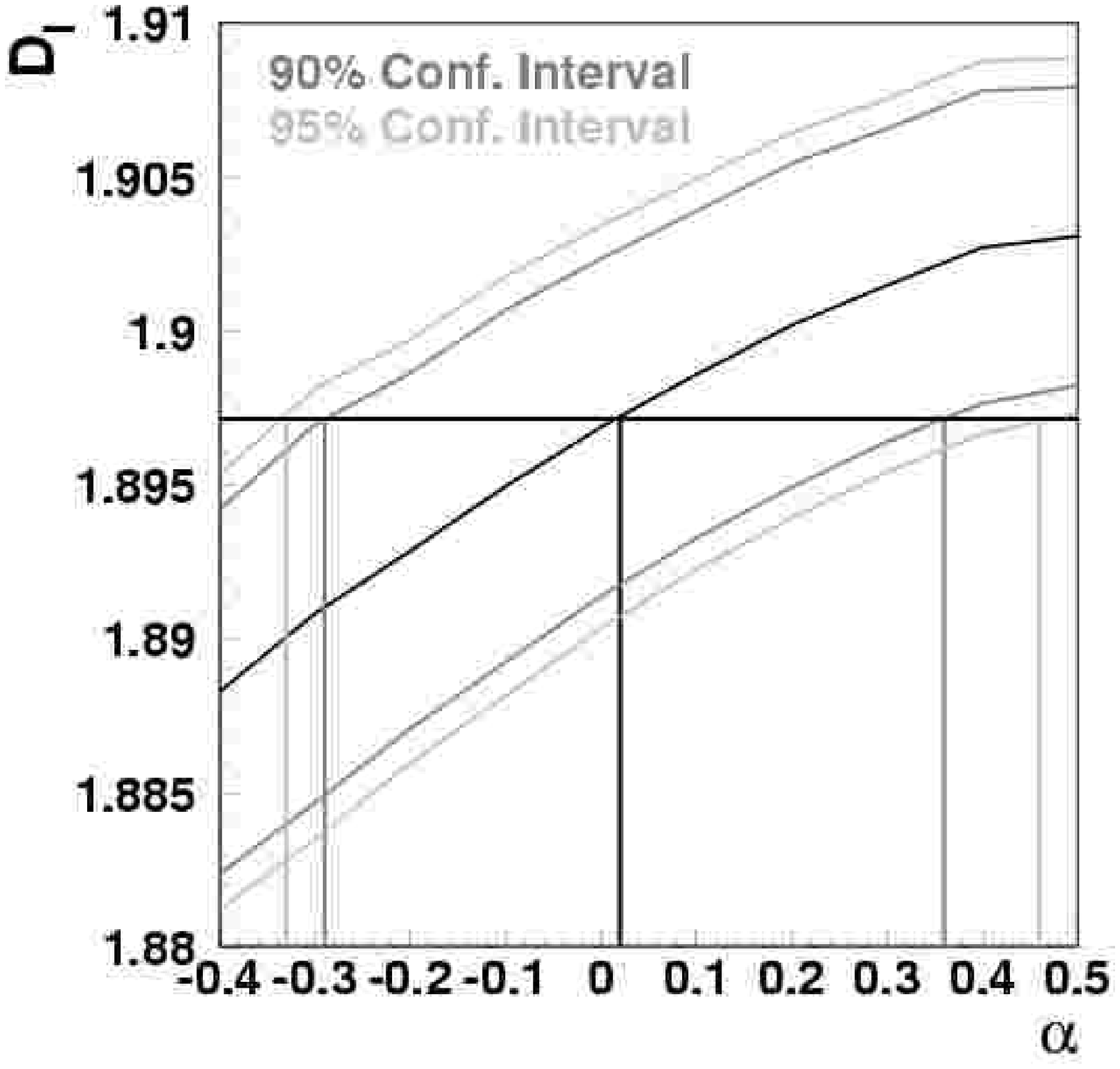}\\
(c)\includegraphics[width=6.15cm]{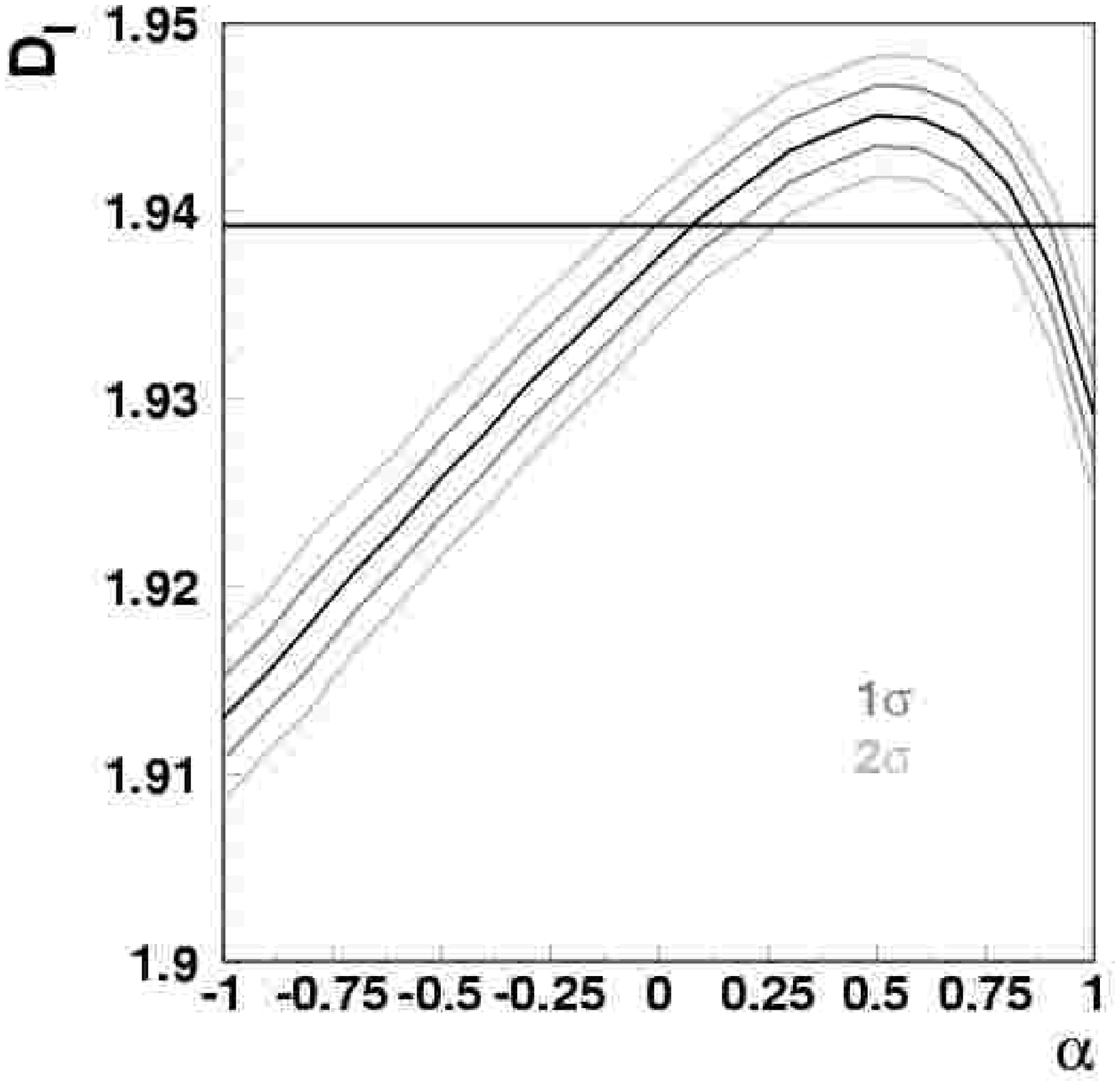}&
(d)\includegraphics[width=6.15cm]{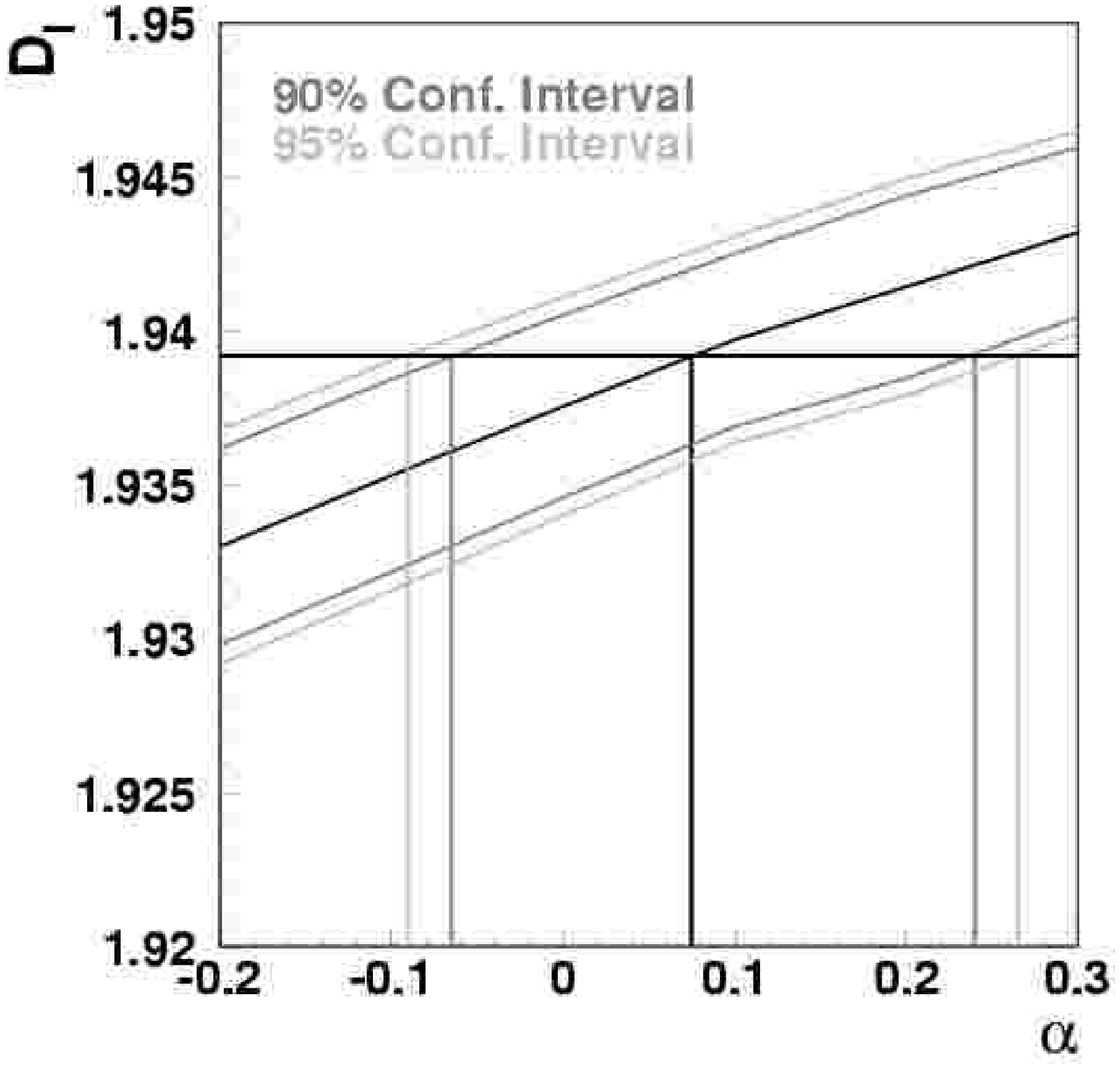}\\
\end{tabular}
\end{center}
\caption{Study of the dependence of $D_{\rm I}$ upon $\alpha$ for a dipole
enhancement source model---(a): $N_{Shower}=500$  $\alpha=.02\pm.21$;
(b): $N_{Shower}=500$ (zoomed); 90\% confidence interval: $[-0.29,0.36]$; 
(c): $N_{Shower}=2000$; $\alpha=.075\pm.085$;
(d): $N_{Shower}=2000$ (zoomed); 90\% confidence interval: $[-0.065,0.24]$.
In each case, the solid solid horizontal line 
indicates the value of $D_{\rm I}$
for the simulated isotropic data set.  The vertical lines in (b) and (d) 
indicate the projection of the nominal value and 90\% and 95\% confidence 
intervals of $\alpha$ on the $x$-axis.}
\label{figure:dipoledi}
\end{figure}

In the case of the dipole enhancement source model in equation 
\ref{equation:dipole}, the scaling parameter is $\alpha$.  
By varying $\alpha$ between $-1$ and $1$, we 
develop a curve which will show the relationship between $D_{\rm I}$ and
$\alpha$.  By considering the actual value of $D_{\rm I}$ for the ``real''
data set, we then establish a nominal value for $\alpha$ and a 90\% 
confidence interval.  The results for both $N_{Shower}=500$ and 
$N_{Shower}=2000$ are shown in figure~\ref{figure:dipoledi}.  In the case of
our simulated isotropic set with $N_{Shower}=500$, 
$\alpha=0.02\pm0.21$ with a 90\% confidence 
interval of $[-0.29,0.36]$.  In the case of our simulated isotropic set with 
$N_{Shower}=2000$, $\alpha=0.075\pm0.085$ with a 90\% confidence 
interval of $[-0.065,0.24]$.  These results tell us that our ``real'' data can,
at the $1\sigma$ level of up to 0.2.  We notice that
in figure~\ref{figure:dipoledi}, there are potentially two suitable intervals 
of $\alpha$ that possess similar values of $D_{\rm I}$.  This emphasizes the 
necessity of making density plots to assure by visual inspection that the
appropriate interval is chosen.  
  
\subsection{Seven Source Model}

\begin{figure}[t,b]
\begin{center}
\begin{tabular}{c@{\hspace{0.0cm}}c}
(a)\includegraphics[width=6.15cm]{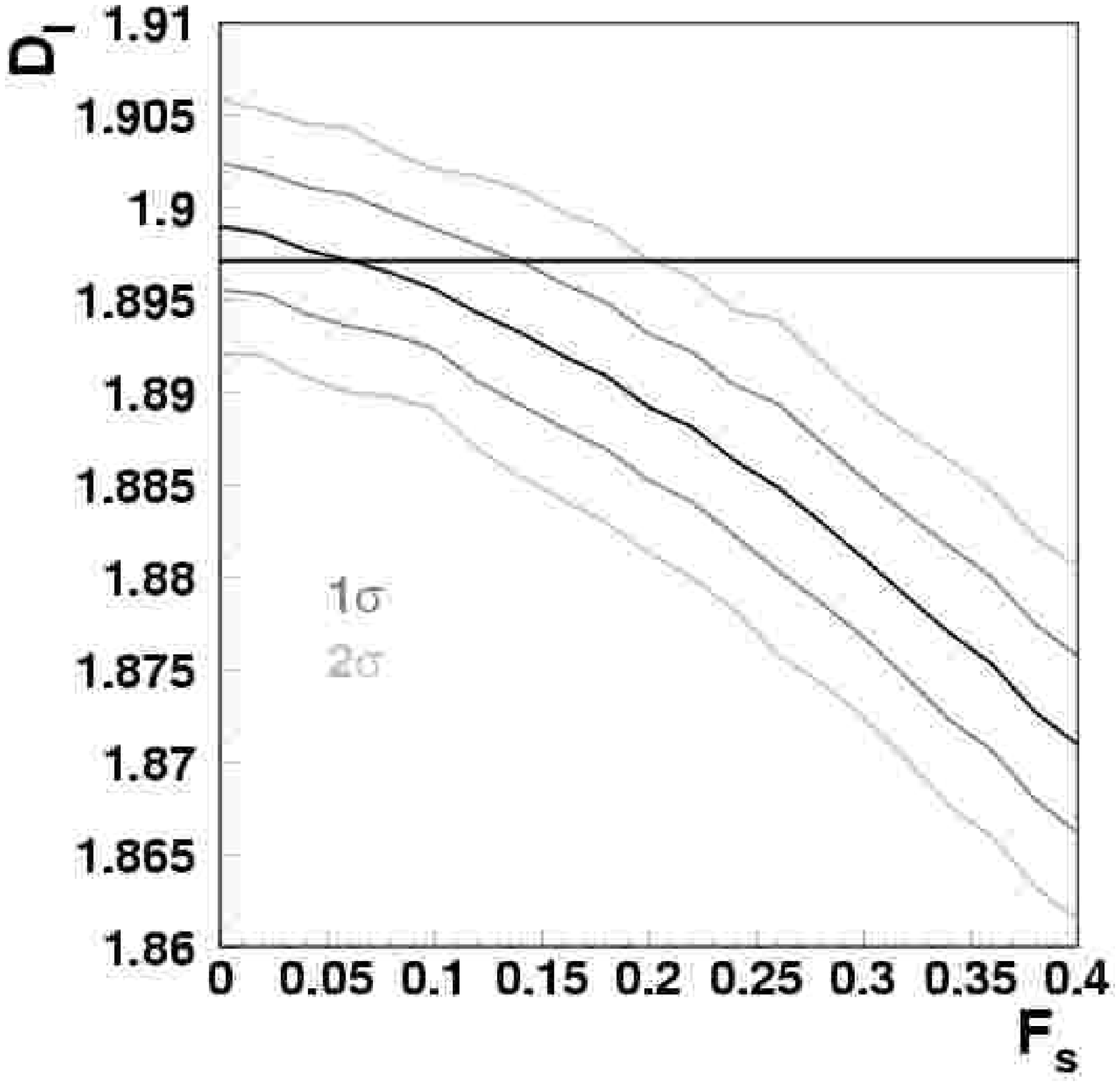}&
(b)\includegraphics[width=6.15cm]{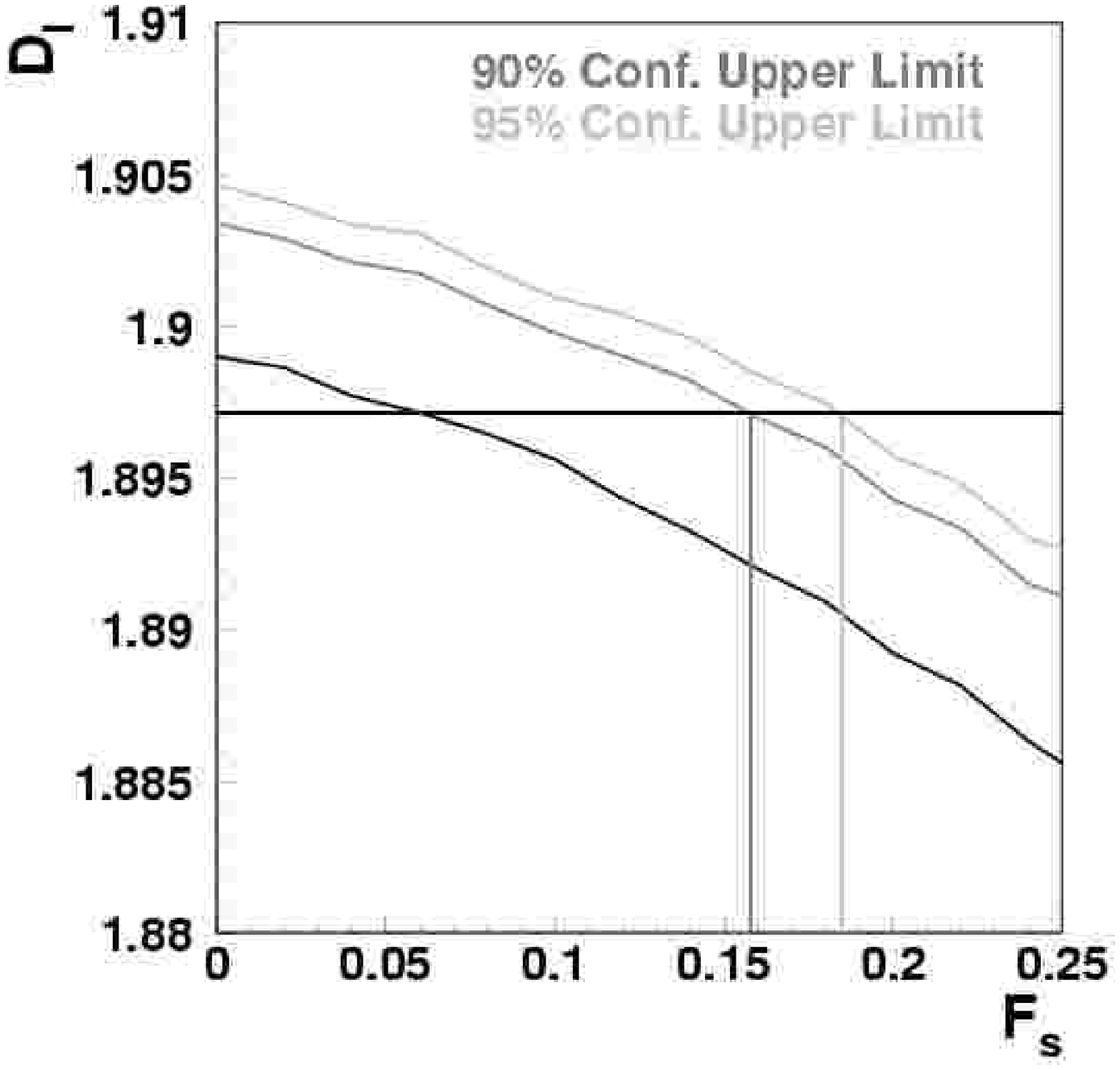}\\
(c)\includegraphics[width=6.15cm]{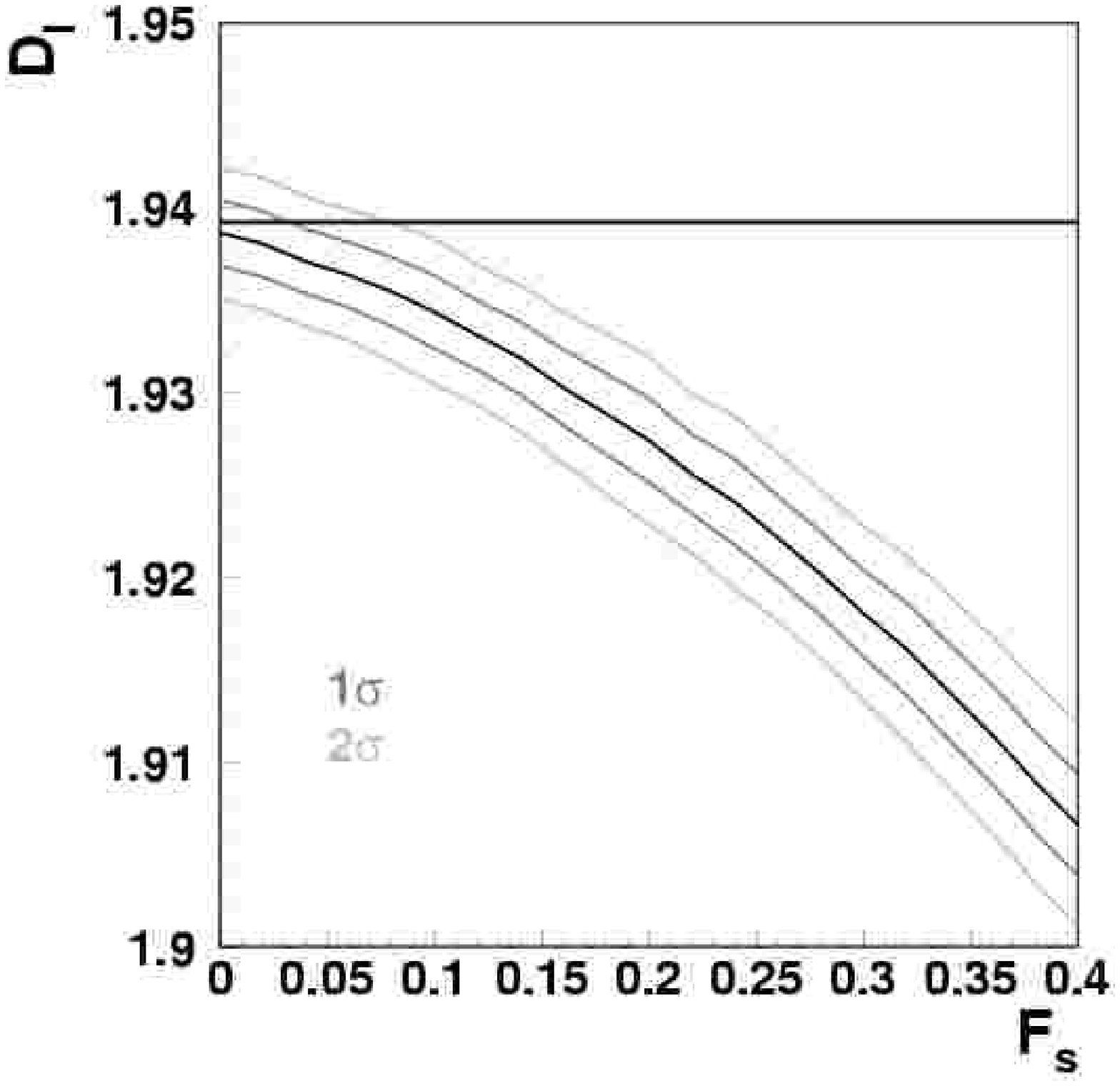}&
(d)\includegraphics[width=6.15cm]{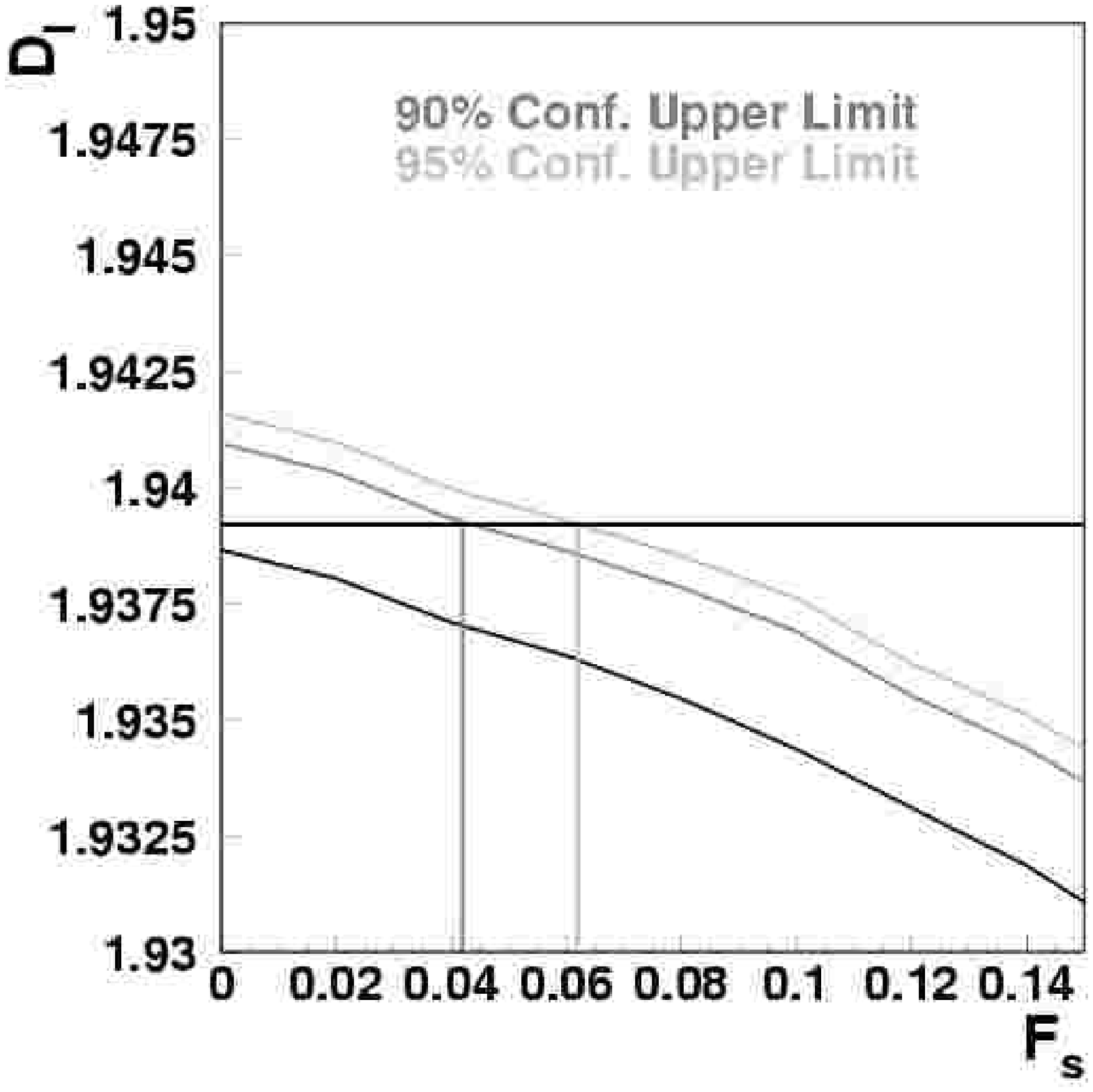}\\
\end{tabular}
\end{center}
\caption{Study of the dependence of $D_{\rm I}$ upon $F_s$ for the 
seven source model---(a): $N_{Shower}=500$;
(b): $N_{Shower}=500$ (zoomed); 90\% confidence upper limit=0.16; 
(c): $N_{Shower}=2000$;
(d): $N_{Shower}=2000$ (zoomed); 90\% confidence upper limite=0.04;.
In each case, the solid solid horizontal line 
indicates the value of $D_{\rm I}$
for the simulated isotropic data set.  The vertical lines in (b) and (d) 
indicate the projection of 90\% and 95\% confidence upper limits
of $r_{\rm s}$ on the $x$-axis.}
\label{figure:7sourcedi}
\end{figure}

In the case of the seven source model, the scaling parameter is 
$F_s$, the fraction of the total event sample 
that is produced by the discrete sources.  
By varying $F_s$ between $0$ and $0.40$, we 
develop a curve which shows the relationship between $D_{\rm I}$ and
$F_s$.  By considering the actual value of $D_{\rm I}$ for the ``real''
data set, we then establish a nominal value for $F_s$ and a 90\% 
confidence upper limit for $F_s$.  
The results for both $N_{Shower}=500$ and 
$N_{Shower}=2000$ are shown in figure~\ref{figure:7sourcedi}.
In the case of our simulated isotropic set with $N_{Shower}=500$, 
we have a 90\% confidence upper limit  
of $F_s=0.16$.  In the case of our simulated isotropic set with 
$N_{Shower}=2000$, we have a 90\% confidence upper limit  
of $F_s=0.04$.  This tells us that our ``real'' data can have, at most, 16\% 
(for 500 events) or 4\% (for 2000 events) of these events coming from the seven
sources.
  
\subsection{Dark Matter Halo Source Model}

\begin{figure}[t,b]
\begin{center}
\begin{tabular}{c@{\hspace{0.0cm}}c}
(a)\includegraphics[width=6.15cm]{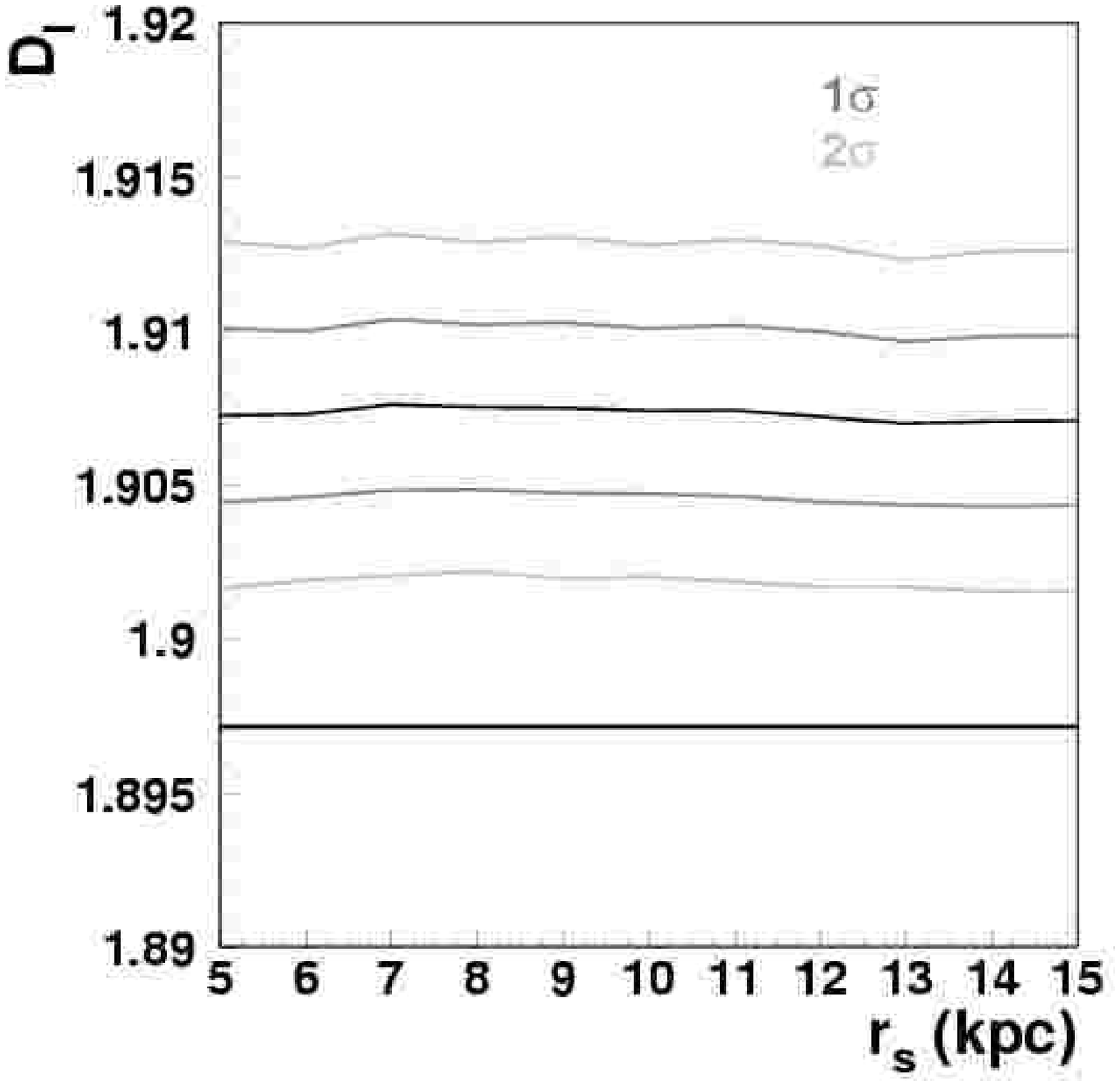}&
(b)\includegraphics[width=6.15cm]{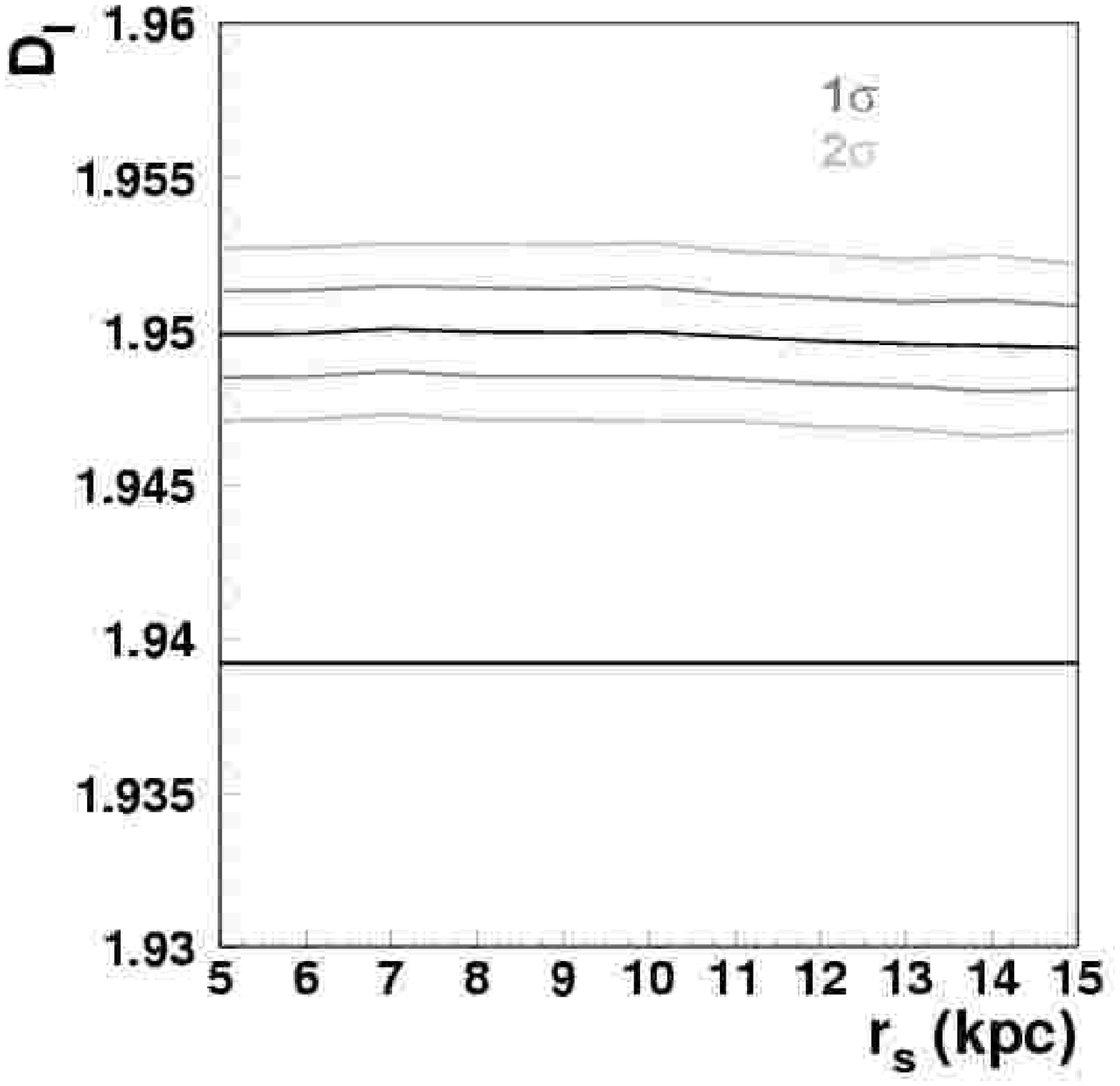}
\end{tabular}
\end{center}
\caption{Comparison of the distribution of $D_{\rm I}$ values for a dark matter
halo source model with a full range of hypothesized values for 
$r_{\rm s}$---(a): $N_{Shower}=500$, the dark matter halo model can be rejected
at a level $\geq3.6\sigma$;
(b): $N_{Shower}=2000$, the dark matter halo model can be rejected
at a level $\geq7.0\sigma$. In each case, the solid horizontal line 
indicates the value of $D_{\rm I}$ for the simulated isotropic data set. }
\label{figure:galhalodi}
\end{figure}

In the case of the dark matter halo source model in equation 
\ref{equation:nfw}, the variable parameter is $r_{\rm s}$, 
the critical radius in the NFW profile \cite{navarro}.  
By varying $r_{\rm s}$ between $5.0$~kpc and
$15.0$~kpc we  
develop the curve which will demonstrate the dependence of $D_{\rm I}$ upon
$r_{\rm s}$.  We can then show that the dark matter halo source model can be
rejected with at level $\geq3.6\sigma$ for $N_{Shower}=500$ and at 
a level of $\geq7.0\sigma$ for $N_{Shower}=2000$ for the full range of 
hypothesized values for $r_{\rm s}$. The results are shown in 
figure~\ref{figure:galhalodi}. 

\section{Discussion}

Fractal dimensionality has several advantages over conventional anisotropy
techniques.  First of all, it naturally accommodates angular resolution.  
This is extremely important when considering event sets with
asymmetric errors, when analysing 
event sets with variable values for the angular 
resolution (e.g. dependent on energy or geometry), or when 
combining multiple data sets from different detectors for a single analysis.

Another advantage that fractal dimensionality possesses is the ability
to accommodate any aperture.  Because this method makes a relative 
comparison between a sample and simulations using the same aperture, the 
physical aperture is simply folded into the analysis.  This once again allows 
the combination of data from multiple detectors with {\it very} different
apertures.  It also allows the analysis of extremely complicated
apertures without the need to include normalizing factors that needlessly 
complicate the predictability of the Poisson fluctuations in the data sample.

Perhaps the most striking feature of the fractal dimensionality 
method is that it only requires a {\it single} 
measurement of one's data.  While fractal dimensionality will 
not always provide better statistical significance 
than a direct measurement for 
a specific anisotropy, the fact that one considers only a single measurement
of the data, for any number of potential anisotropic models, provides one the
means to simultaneously reject or accept all of those models without the 
ensuing statistical penalty.  

A possible way of increasing the sensitivity of the fractal dimensionality
method is by 
considering the general case of case of $D_q$.  By varying the value of $q$
to something other than $1$, it might be possible to increase the sensitivity 
of this method to various anisotropies. 

The fractal dimensionality method does have some weaknesses.  First and 
foremost is the potential for multiple solutions as was demonstrated in the
dipole source model above.  The method cannot be applied blindly.
It requires a careful inspection of both the data sample and the
simulated samples in order to resolve possible ambiguities.  Another drawback
is the amount of computation required to calculate 
$D_{\rm I}$ for a large number of simulated data sets.  Producing just
the plots in figure~\ref{figure:dipoledi}c consumed the equivalent
of $\sim1000$~CPU hours on 1GHz Athlon machine.  
Another limitation is the potential 
for different source models to effectively cancel each other out and yield
a potentially deceiving value of $D_{\rm I}$ 
that resembles that of an isotropic source. One solution 
for this is to separately consider the value of $D_{\rm I}$ 
for different celestial regions in the data.  Of course, this will 
incur a statistical penalty. 

Two particularly useful roles for fractal dimensionality analysis are (a): as
a first test to ascertain if a sample possesses the same heterogeneity as
the expected isotropic background and (b): as a wholly independent
observable that can be used to confirm the results of a direct measurement.

Today the HiRes detector continues to accumulate data.  Soon the Auger
detector will be acquiring data at seven times the rate of HiRes.
As the number of observed UHECRs continues to rise, fractal dimensionality
will grow ever more effective in its ability to discern the between
potential source models in the continuing effort to solve the mystery 
of UHECRs.  
  
\section{Acknowledgments}
This work is supported by US NSF grants PHY-9321949 PHY-9974537, and
PHY-0140688. We would like to thank Professors Glennys Farrar and Benjamin Bromley
for invaluable insights that went into the analysis used to produce this paper.
We gratefully acknowledge the contributions from the University
of Utah Utah Center for High Performance Computing.

\end{document}